\documentclass[12pt,preprint]{aastex}

\received{}
\accepted{}

\author{
Gordon T. Richards,\altaffilmark{1}
Robert C. Nichol,\altaffilmark{2}
Alexander G. Gray,\altaffilmark{3}
Robert J. Brunner,\altaffilmark{4}
Robert H. Lupton,\altaffilmark{1}
Daniel E. Vanden Berk,\altaffilmark{5}
Shang Shan Chong,\altaffilmark{2}
Michael A. Weinstein,\altaffilmark{6,7}
Donald P. Schneider,\altaffilmark{6}
Scott F. Anderson,\altaffilmark{8}
Jeffrey A. Munn,\altaffilmark{9}
Hugh C. Harris,\altaffilmark{9}
Michael A. Strauss,\altaffilmark{1}
Xiaohui Fan,\altaffilmark{10}
James E. Gunn,\altaffilmark{1}
\v{Z}eljko Ivezi\'{c},\altaffilmark{1,8}
Donald G. York,\altaffilmark{11,12}
and J. Brinkmann\altaffilmark{13}
}

\altaffiltext{1}{Princeton University Observatory, Peyton Hall, Princeton, NJ 08544.}
\altaffiltext{2}{Department of Physics, Carnegie Mellon University, 5000 Forbes Avenue, Pittsburgh, PA 15232.}
\altaffiltext{3}{Robotics Institute, Carnegie Mellon University, 3128 Newell-Simon Hall, 5000 Forbes Avenue, Pittsburgh, PA 15213-3891.}
\altaffiltext{4}{Department of Astronomy, University of Illinois at Urbana-Champaign, 1002 West Green Street, Urbana, IL 61801-3080.}
\altaffiltext{5}{Department of Physics and Astronomy, University of Pittsburgh, 3941 O'Hara Street, Pittsburgh, PA 15260.}
\altaffiltext{6}{Department of Astronomy and Astrophysics, The Pennsylvania State University, 525 Davey Laboratory, University Park, PA 16802.}
\altaffiltext{7}{Department of Physics, Astronomy, and Geophysics, Connecticut College, Box 5622, 270 Mohegan Avenue, New London, CT, 06320.}
\altaffiltext{8}{Department of Astronomy, University of Washington, Box 351580, Seattle, WA 98195.}
\altaffiltext{9}{US Naval Observatory, Flagstaff Station, P.O. Box 1149, Flagstaff, AZ 86002-1149.}
\altaffiltext{10}{Steward Observatory, University of Arizona, 933 North Cherry Avenue, Tucson, AZ 85721.}
\altaffiltext{11}{Department of Astronomy and Astrophysics, The University of Chicago, 5640 South Ellis Avenue, Chicago, IL 60637.}
\altaffiltext{12}{Enrico Fermi Institute, The University of Chicago, 5640 South Ellis Avenue, Chicago, IL 60637.}
\altaffiltext{13}{Apache Point Observatory, P.O. Box 59, Sunspot, NM 88349.}

\begin{document}

\title{Efficient Photometric Selection of Quasars from the Sloan
Digital Sky Survey: 100,000 $z<3$ Quasars from Data Release One}

\begin{abstract}

We present a catalog of 100,563 unresolved, UV-excess (UVX) quasar
candidates to $g=21$ from 2099 deg$^2$ of the Sloan Digital Sky Survey
(SDSS) Data Release One (DR1) imaging data.  Existing spectra of
22,737 sources reveals that 22,191 (97.6\%) are quasars; accounting
for the magnitude dependence of this efficiency, we estimate that
95,502 (95.0\%) of the objects in the catalog are quasars.  Such a
high efficiency is unprecedented in broad-band surveys of quasars.
This ``proof-of-concept'' sample is designed to be maximally
efficient, but still has 94.7\% completeness to unresolved, $g\lesssim
19.5$, UVX quasars from the DR1 quasar catalog.  This efficient and
complete selection is the result of our application of a probability
density type analysis to training sets that describe the 4-D color
distribution of stars and spectroscopically confirmed quasars in the
SDSS.  Specifically, we use a non-parametric Bayesian classification,
based on kernel density estimation, to parameterize the color
distribution of astronomical sources --- allowing for fast and robust
classification.  We further supplement the catalog by providing
photometric redshifts and matches to FIRST/VLA, ROSAT, and USNO-B
sources.  Future work needed to extend the this selection algorithm to
larger redshifts, fainter magnitudes, and resolved sources is
discussed.  Finally, we examine some science applications of the
catalog, particularly a tentative quasar number counts distribution
covering the largest range in magnitude ($14.2<g<21.0$) ever made
within the framework of a single quasar survey.

\end{abstract}

\section{Introduction}

Since the discovery of quasars \markcite{sch63}({Schmidt} 1963), ambitious surveys
\markcite{sg83,fch+87,bsc+00,yaa+00}(e.g., {Schmidt} \& {Green} 1983; {Foltz} {et~al.} 1987; {Boyle} {et~al.} 2000; {York} {et~al.} 2000) have caused the number of
known quasars to rise from one to tens of thousands.  Yet even in this
day of very large surveys and deep digital imaging, we are still far
from identifying the more than 1.6 {\em million} $z<3$ quasars that
are expected to fill the celestial sphere to $g\sim21$.  The problem
lies not in covering enough of the sky to faint enough magnitudes, but
rather in the efficient separation of quasars from other astronomical
sources.  Current algorithms are typically more than 60\% efficient
for UV-excess (UVX) quasars to relatively bright magnitudes, but the
selection efficiency drops toward fainter magnitudes where the
photometric errors are largest and most of the observable objects
reside.  Further complicating the issue is the need to obtain spectra
for each candidate.\footnote{X-ray to optical flux ratios may also
suffice, but X-ray detections can take just as long to obtain.}  Thus
surveys of quasars would benefit considerably from algorithms with
selection efficiencies that mitigate the need for confirming spectra.
We describe such an algorithm based on the photometric data of the
Sloan Digital Sky Survey (SDSS; \markcite{yaa+00}{York} {et~al.} 2000).

Optical surveys for quasars, including the SDSS, typically rely on
simple color cuts in two or more colors to select objects that are
likely to be quasars and to reject objects that are unlikely to be
quasars.  The color selection part of the SDSS's quasar algorithm
\markcite{rfn+02}({Richards} {et~al.} 2002) is essentially two, 3-D
color selection algorithms.  One branch of the algorithm uses the
$ugri$ bands to identify UVX quasars, the other uses the $griz$ bands
to identify $z>3$ quasars.  

Another way to select quasars from imaging data is to use known
quasars to determine what regions of color space quasars occupy.  Once
these regions have been identified, spectroscopic quasar target
selection involves simply observing objects from those regions of
color space that are most likely to yield quasars (or perhaps least
likely to yield significant number of contaminants).  At the beginning
of the SDSS survey, the construction of such an algorithm would have
been difficult given the lack of data, but with the current abundance
of SDSS imaging data and spectroscopic follow-up, it is now possible
to design such algorithms.

The approach used in this paper is based on the simple philosophy that
the most efficient and complete way to find quasars is to target those
regions of color space dominated by quasars and/or that have
sufficiently low rates of contamination that we can afford to probe
them for quasars.  To accomplish this goal, we take advantage of an
existing statistical technique known as kernel density estimation
(KDE; \markcite{sil86,gra04} Silverman 1986; {Gray} {et al.} 2004).  By
applying this technique to ``training sets'' of stars and quasars we
can optimally classify ``test sets'' of potential quasar candidates.
Our algorithm takes advantage of and goes well beyond the
color-characterization of small samples of quasars
\markcite{rfs+01}(e.g., {Richards} {et~al.} 2001a), to efficiently
select quasars (or other objects for that matter) from much larger
samples of 5-band SDSS imaging data.  In fact, the algorithm is so
efficient that the failure rate is comparable to that of automated
identification of quasar {\em spectra} from the SDSS, and thus, for
many science applications, spectroscopy is {\em not} needed.
Furthermore, we capitalize on the structure in the quasar
color-redshift relation to compute relatively accurate photometric
redshifts \markcite{rws+01,bcs+01,wrs+04}({Richards} {et~al.} 2001b;
{Budav{\' a}ri} {et~al.} 2001; {Weinstein et al.} 2004) for all the
resulting quasar candidates.  The end product is a catalog of over
100,000 $z<3$ quasar candidates with photometric redshifts that were
selected from the 2099 deg$^2$ of SDSS DR1 imaging data.

In \S~2 we describe our input data.  Section~3 presents an overview of
the algorithm and the details of its application.  In \S~4 we present
the catalog and discuss its completeness and efficiency along with
matching to other catalogs and the computation of photometric
redshifts.  Section~5 presents some ideas for future improvement and
\S~6 discusses some science highlights.

\section{The SDSS DR1 Imaging Data}

The imaging data that was used as the basis for our catalog is
contained in the SDSS First Data Release (DR1; \markcite{aaa+03}{Abazajian} {et~al.} 2003).  The
SDSS is a project to map roughly 10,000 deg$^{2}$ of sky in 5
photometric passbands ($ugriz$) using a large-format CCD camera
\markcite{gcr+98}({Gunn} {et~al.} 1998).  The characterization of the photometric system is
discussed by \markcite{fig+96}{Fukugita} {et~al.} (1996), \markcite{hfs+01}{Hogg} {et~al.} (2001), \markcite{stk+02}{Smith} {et~al.} (2002), and
\markcite{slb+02}{Stoughton} {et~al.} (2002).  Unless otherwise stated, all magnitudes discussed
herein are ``asinh'' point-spread-function magnitudes \markcite{lgs+99}({Lupton}, {Gunn}, \& {Szalay} 1999)
that have been dereddened according to \markcite{sfd98}{Schlegel}, {Finkbeiner}, \&  {Davis} (1998).  The astrometric
accuracy of the survey (and thus of the catalog presented herein) is
better than 100 mas per coordinate rms \markcite{pmh+03}({Pier} {et~al.} 2003).  Our work
further makes use of the spectroscopic follow-up of quasars; the
selection algorithm is described by \markcite{rfn+02}{Richards} {et~al.} (2002), the tiling
algorithm by \markcite{blm+03}{Blanton} {et~al.} (2003), and the DR1 spectroscopic quasar catalog
by \markcite{sfh+03}{Schneider} {et~al.} (2003).  In \S\S~\ref{sec:train} and \ref{sec:test} we
describe the cuts that were applied to the imaging data to construct
our training and test sets.

\section{Nonparametric Bayesian Classification}
\subsection{Overview of the Algorithm}

The basic idea of our quasar selection algorithm is conceptually as
follows.  We wish to classify a set of unlabeled objects (the {\it
test set}) as either stars or quasars.  We first create samples of
``stars'' and ``quasars'' that will serve as {\em training sets}.  For
each object in our test set that we wish to classify, we compute its
probability of being a star and its probability of being a quasar.
The test object is assigned the label corresponding to the higher
probability.

The ``probability of being a star'' for an object $x$ (represented by
four color measurements) can be formalized as the likelihood of $x$
under the probability density function (pdf) which describes stars,
i.e., $p(x|C_1)$, where $C_1$, or class 1, is the star class.  This
pdf could be represented, for example, as a histogram.  Because our
measurement space consists of four color dimensions, this would
correspond to a 4-D grid of counts.  Instead, we will use a {\it
kernel density estimate} (KDE; \markcite{sil86}Silverman 1986) of the
pdf of stars.  This mature statistical method is a powerful
generalization of the concept of a histogram which yields a more
accurate estimate of the true underlying pdf.  Instead of discrete
bins whose locations are defined by a grid, KDE defines each `bin' by
its center point and the extent of the bin by a continuous {\it kernel
function} --- for example a Gaussian function in 4-D.  We describe KDE
in more detail later in the paper.  For an introduction to density
estimation, we refer the reader to \markcite{sil86} Silverman (1986).

Once we have a way of estimating the likelihood of $x$ being a star
(or quasar), or the value at $x$ of the star pdf (or quasar pdf), we
could simply choose the class corresponding to the higher likelihood.
However, we will incorporate one further piece of information before
determining the ``probability of being a star'' --- the user's prior
belief that the object is a star, denoted $P(C_1)$. This captures any
and all subjective information which the user may have outside of
observing the actual training set --- namely here, the fraction of an
unseen set of objects which the user roughly expects to be stars.  To
incorporate this prior information with the likelihoods given by KDE,
we use a simple application of Bayes' Rule
\markcite{bayes1763,ptv+92}(Bayes 1763; {Press} {et~al.} 1992), which
weights each likelihood with its corresponding prior probability to
obtain the a posterior probability of being a star or quasar:

\begin{equation}
P(C_1|x) = \frac{ p(x|C_1) P(C_1) }{ p(x|C_1) P(C_1) + p(x|C_2) P(C_2)}
\end{equation}

Specifically, in our context, objects with $P(C_1|x)>0.5$ are
classified as stars, while objects with $P(C_1|x)<0.5$ are classified
as quasars.  We refer to the resulting overall classifier as a
nonparametric Bayes classifier (NBC), for lack of a standard name in
the statistical literature.\footnote{This is sometimes also called
kernel discriminant analysis or kernel density classification.}

\subsection{The Training Sets}
\label{sec:train}

For the quasar training set, we simply used the four primary SDSS
colors ($u-g$, $g-r$, $r-i$, $i-z$) of the 16,713 quasars from
\markcite{sfh+03}{Schneider} {et~al.} (2003) without applying any additional cuts based on
luminosity, morphology, method of selection, photometric errors, etc.
These quasars span redshift and magnitude ranges of $0.08 \le z \le
5.4$ and $14.99 \le i \le 21.55$.

For the stars training set, we used the four primary SDSS colors
extracted from a random sample of 10\% of all point
sources\footnote{We define point sources as those that have {\tt
objc\_type$\,=6$} in the SDSS's photometric database.  Morphologic
classification accuracy is a function of magnitude, being nearly
perfect for $g<20$, but only 90\% accurate at our $g=21$ limit.} in
the DR1 imaging area with $14.5<g<21.0$.  We rejected any objects that
did not pass the photometric quality tests that the SDSS quasar
algorithm applies before it does color-selection.  Specifically, we
rejected those objects that failed the ``fatal'' or ``non-fatal''
error tests; see \markcite{rfn+02}{Richards} {et~al.} (2002) for details.  Finally, since we have
included {\em all} point sources --- including quasars --- in the
stars training set, we have also rejected any spectroscopically
confirmed quasars and any radio sources (which, for unresolved
sources, are more likely to be quasars than anything else).  The total
number of objects in the initial stars training set was 478,144.  The
final stars training set that we used to classify our objects went
through an additional ``cleaning'' pass described in
\S~\ref{sec:clean}.

\subsection{The Test Set}
\label{sec:test}

The goal of this paper is to present a ``proof-of-concept'' of the NBC
approach to efficient selection of astronomical objects.  Thus we
start where quasar selection is admittedly easiest.  The test set for
which we have determined star/quasar classifications consists of
SDSS-DR1 point sources with $u-g<1.0$, i.e., UVX sources that were
selected from the {\tt PhotoPrimary} DR1 database table.  Currently we
exclude sources that are resolved in the SDSS imaging data.  We
further restrict the sample to those objects with observed $g$
magnitudes fainter than 14.5\footnote{We use observed (i.e., not
dereddened) magnitudes for the bright limit since the purpose of this
limit is to reject objects that may be saturated in the imaging.},
dereddened $g$ magnitudes brighter than 21.0, and $u$-band errors less
than 0.5 mag (i.e., are at least $2\sigma$ detections in $u$).  We use
the $g$ band instead of $i$ since our selection is a UVX one, meaning
that our quasar candidates will generally have $z<3$ and little
$g$-band Ly$\alpha$ forest absorption; this choice also facilitates
comparison with previous work such as the 2dF QSO Redshift Survey
(2QZ; \markcite{csb+04}{Croom} {et~al.} 2004).  As with the stars
training set above, we reject objects that fail the ``fatal'' or
``non-fatal'' error tests used by the official SDSS quasar selection
algorithm \markcite{rfn+02}({Richards} {et~al.}  2002).  The full test
set contains $831,600$ objects.

\subsection{Kernel Density Estimation}

Once we have defined the training and test sets we can begin the
process of classification by computing the likelihood of each object
$x$ in the test set with respect to each training set (or
equivalently, the density at $x$ under the stars and under the
quasars), using the nonparametric (i.e., distribution-free) {\em
kernel density estimator}:
\begin{equation}
\hat{p}(x) = \frac{1}{N} \sum_i^N K_h(||x - x_i||)
\end{equation}
where $N$ is the number of data points, $K_h(z)$ is called the kernel
function and satisfies $\int_{-\infty}^{\infty} K_h(z) dz = 1$, $h$ is
a scaling factor called the bandwidth, and $z$ is the ``distance''
between a point in the test set to a point in the training set (in our
case, these distances are 4-D Euclidean color differences, $||x -
x_i||$).  In this work we mainly use a Gaussian kernel, where the
bandwidth corresponds to the variance of the Gaussian, i.e. $K_h(||x -
x_i||) = \frac{1}{h\sqrt{2\pi}}\exp{-\frac{||x - x_i||^2}{2h^2}}$.  To
make an analogy with a 1D histogram (the simplest kind of density
estimator), the reader can think of a 1D histogram with a bin width of
$h$ as being a kernel density estimator with $K_h(z) = {\rm constant}$
and with the location of the bins being fixed rather than being
centered at the individual data points.

Kernel density estimation is the most widely-used and well-studied
method for nonparametric density estimation, owing to both its
simplicity and flexibility, and the many theorems establishing its
consistency for near-arbitrary unknown densities and rates of
convergence for its many variants.  See \markcite{sil86} Silverman
(1986) for more details regarding the concept of kernel density
estimation.

\subsection{Fast Algorithms}

A naive algorithm for computing the kernel density estimate at one
point among $N$ points requires $N$ distance operations.  Computing
the density estimate at $N$ points among $N$ points thus scales as
$N^2$.  This is intractable for large datasets such as ours.  We use a
fast computational algorithm based on space-partitioning trees and
principles similar to those used in $N$-body solvers \markcite{gm03}({Gray} \& {Moore} 2003).  We
will refer to this as the ``fast KDE'' algorithm.

For the work in this paper a second, new algorithm was developed, for
the different computational problem of {\it quickly finding the higher
posterior probability}, i.e., finding the label for each test point
more quickly than by explicitly finding its density under each of the
two class training sets.  A detailed description of the modifications
of the standard NBC algorithm to make it faster are beyond the scope
of this paper and will be described in a future publication
\markcite{gra04}({Gray et al.} 2004). We will refer to this algorithm
as ``fast NBC''.

The fast NBC algorithm need not estimate the density completely for
each object to be classified.  The algorithm need only maintain upper
and lower bounds on the density for each class; the code stops
considering additional data when it finds that the bounds no longer
overlap.  It is exact, i.e., computes the classification labels as if
the kernel density estimates had been computed exactly.  For
additional speed we use an Epanechnikov kernel for this step, which is
Gaussian-like but has finite rather than infinite extent.  The
resulting bandwidths are then scaled appropriately to find the optimal
bandwidth for a true Gaussian kernel.\footnote{For the Epanechnikov
kernel, the bandwidth specifies its entire extent; for the Gaussian
kernel, it specifies one standard deviation --- about $1/3$ of the
entire coverage.}

\subsection{Bandwidth Determination}

The critical step in the KDE process is determining the optimal
``bandwidths'' for kernel density estimation, i.e., the bandwidth that
spans the color space of each training set most efficiently.  This
process is similar to that of deciding upon the best bin size to
represent data in a histogram; using too small of a bin can cause
artificial spikes in the histogram due to small number statistics,
whereas using too large of a bin can hide real information
\markcite{sil86} (Silverman 1986).  There exist mature algorithms for
choosing the bandwidth for KDE which minimize a statistical measure of
the difference between the true underlying pdf and the estimated pdf.
Perhaps the most accepted method for performing this is {\em least
squares cross-validation} \markcite{sil86}(Silverman 1986).  Initially
we used this method to automatically determine the optimal bandwidths
for the two classes separately.  However, with a half million objects
in the stars training set, this method of computing the (optimal)
bandwidth was too computationally intensive.  Furthermore, this
approach corresponds to NBC where the priors are equal, and where the
bandwidths for the stars and quasars are estimated {\em
independently}, based on a statistical criterion (leave-one-out,
cross-validated, least-squares error) for optimal {\em density
estimation} for each class.

Instead, we chose the bandwidth pair using a statistical criterion for
optimal {\em classification accuracy} (leave-one-out, cross-validated
accuracy, in this case.)  This has significant advantages over the
previous method.  First, it considers the bandwidths for both training
sets simultaneously, as a pair, rather than independently.  Optimal
bandwidth selection for density estimation is known to be a difficult
statistical problem.  Estimating each bandwidth independently
compounds the problem that the true criterion of interest is the
behavior of both bandwidths in unison, in terms of the performance of
the resulting classifier; estimating parameters for a classifier is
known to be a statistically easier problem (for example in terms of
convergence rate).

Secondly, the density estimation approach is inherently more
computationally difficult.  As noted above, the fast KDE algorithm
\markcite{gm03}({Gray} \& {Moore} 2003) must inherently perform more
work than the fast NBC algorithm used here \markcite{gra04}({Gray et
al.} 2004).  Using the fast NBC algorithm allows quick computation of
the leave-one-out accuracy score for each pair of bandwidths for our
two training sets.

In the context of leave-one-out accuracy, one ideally determines the
bandwidths by maximizing the classification accuracy of each training
set simultaneously.  However, in our case, we fully expect the
algorithm to misclassify some quasars as stars.  For example a small
fraction of quasars are known to be dust reddened and are more likely
than unreddened quasars to have colors more similar to stars.
Similarly, we are aware that the definition of our star training set
is not exclusive to stars.  Thus we chose to maximize the
classification accuracy after first accounting for reasonable
expectations for misclassification.  The best resulting bandwidth was
0.15 mag for each of the training sets, which resulted in an accuracy
of 94.48\% for the quasars and 97.91\% for the stars.

\subsection{Cleaning the Stars Training Set}
\label{sec:clean}

Once we have bandwidths for each of our training sets, we can simply
apply them to our test set to classify our objects.  However, before
final classification of our test set, we first chose to clean the
stars training set by running it through the algorithm as a test set
since our ``stars'' training set really consists of objects that are
known only to be point sources.  Therefore, we have removed from the
stars training set any objects that were classified by NBC as quasars.
The final ``cleaned'' training set of stars contained $468,149$
objects.  This process is admittedly somewhat circular, but is
appropriate for the goal of this paper, which is to produce a sample
that is as efficient as possible, leaving improvements in completeness
to later work.  After having thus cleaned the stars training set, the
same $0.15$ mag bandwidths resulted in an accuracy within the training
sets of 95.86\% for the quasars and 99.89\% for the stars.

\subsection{Application to the Test Set via NBC}

Once the two bandwidth parameters (from the quasars and cleaned stars
training sets) are finalized, we proceed with the classification of
the objects in the test set, by computing $P(C_1|x)$ for each test
object $x$.  We use the Bayesian prior $P(C_1)$ = 0.88.  This is based
on the fraction of objects in our test set that we believe are likely
to be stars (88\%) given previous testing of the algorithm.  The NBC
classification of our $831,600$ UVX point source objects resulted in
$113,674$ (13.7\%) objects classified as quasars and $717,926$
(86.3\%) classified as stars.

The color distribution of these $113,674$ does indeed strongly
resemble that of the input quasar training set.  However, it was
obvious that there was still a considerable amount of contamination,
primarily faint F-stars, which have errors and metallicities that push
them well into the usual quasar locus.  Since our current algorithm
considers only the colors and not magnitudes (see
\S~\ref{sec:future}), these objects are difficult to remove with the
NBC algorithm and a single prior.

Thus, the initial classification was supplemented by going back and
computing the full KDE star/quasar densities for each of the $113,674$
objects that were classified as quasars.  In this process we used a
Gaussian rather than an Epanechnikov kernel and bandwidths $1/3$ the
size of the above (see above), which yields specific quasar/star
densities for each object as opposed to simple binary classification.
This calculation is now feasible since the number of objects under
consideration has been reduced from the original $831,600$ to
$113,674$ and we have already decided on a bandwidth to use.

\begin{figure}[p]
\epsscale{1.0}
\plottwo{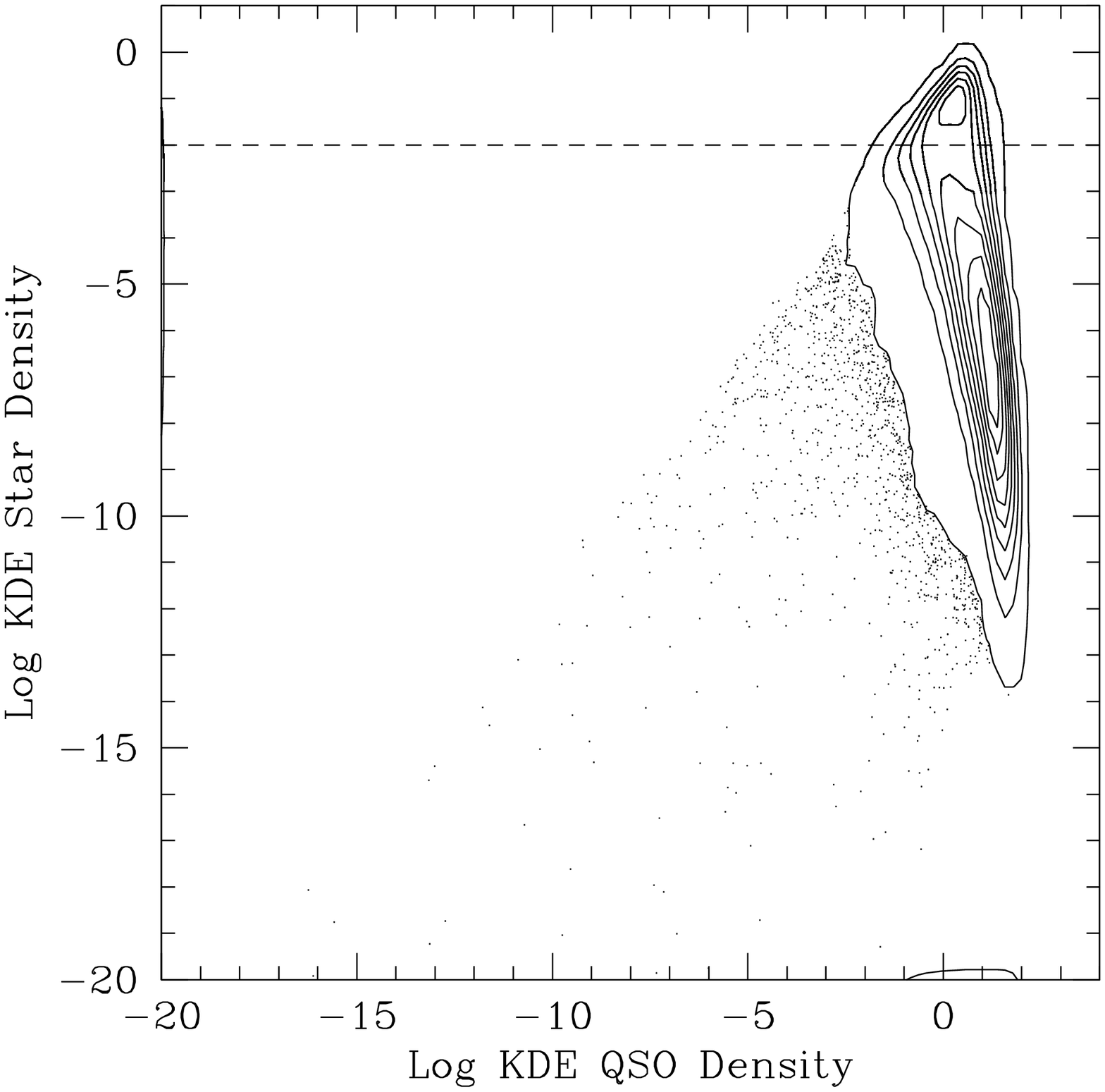}{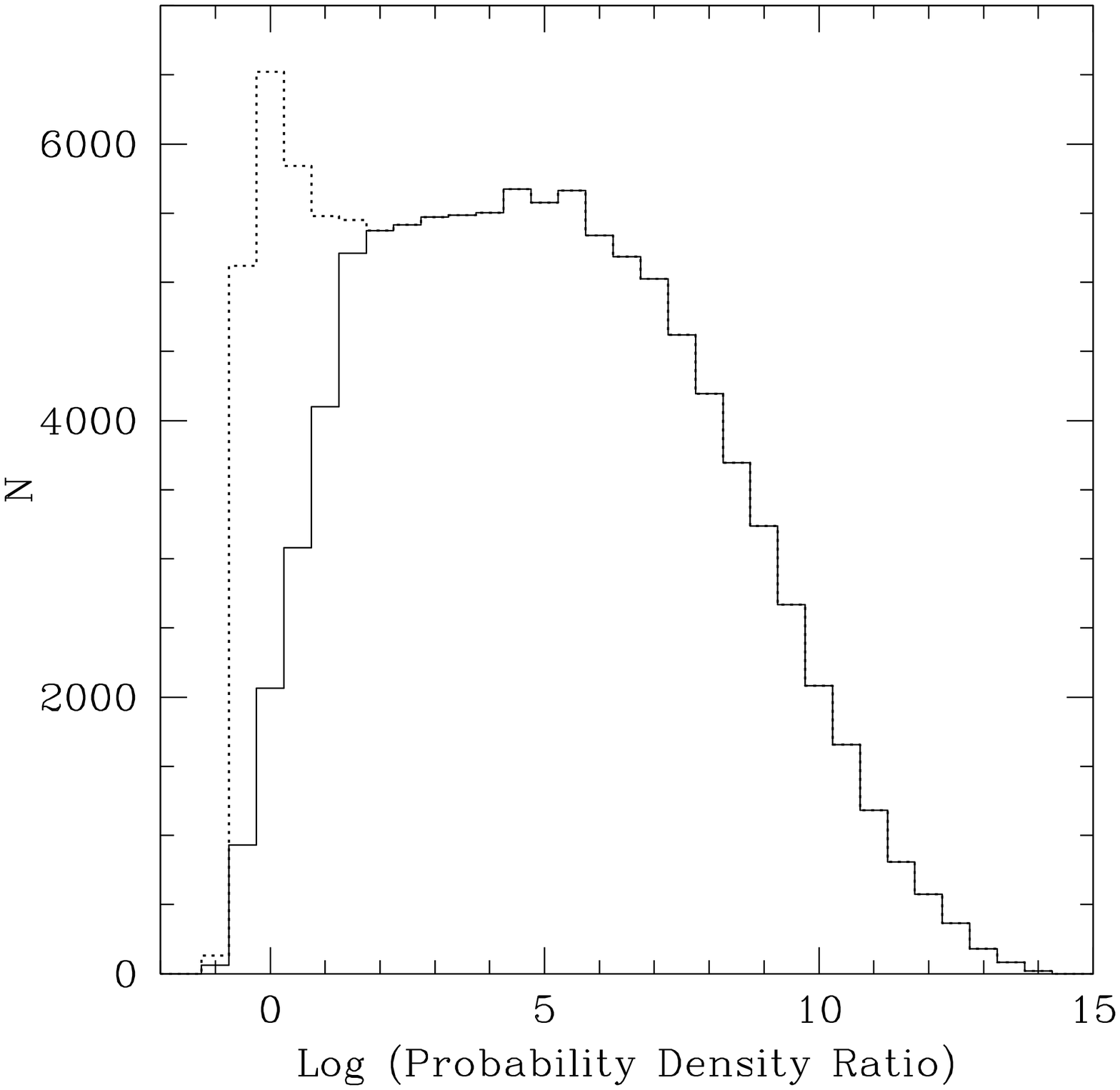}
\caption{{\em Left:} Log of the KDE quasar vs.\ star density.  Larger
values indicate greater likelihood of class membership.  The lack of
objects in the upper-left is a result of the NBC classification that
was used to perform the initial rejection of objects classified as
stars.  The dashed line shows the additional cut that was used to
reject stellar contaminants from the NBC-classified quasars. {\em
Right:} The distribution of the logarithm of the quasar density
divided by the star density.  Larger values indicate greater
probability of membership in the quasar class.  All NBC-classified
quasars are given by the {\em dotted line} and quasars meeting the
additional cut on stellar density (dashed line in left panel) are
given by the {\em solid line}.
\label{fig:fig1}}
\end{figure}

\begin{figure}[p]
\epsscale{1.0}
\plotone{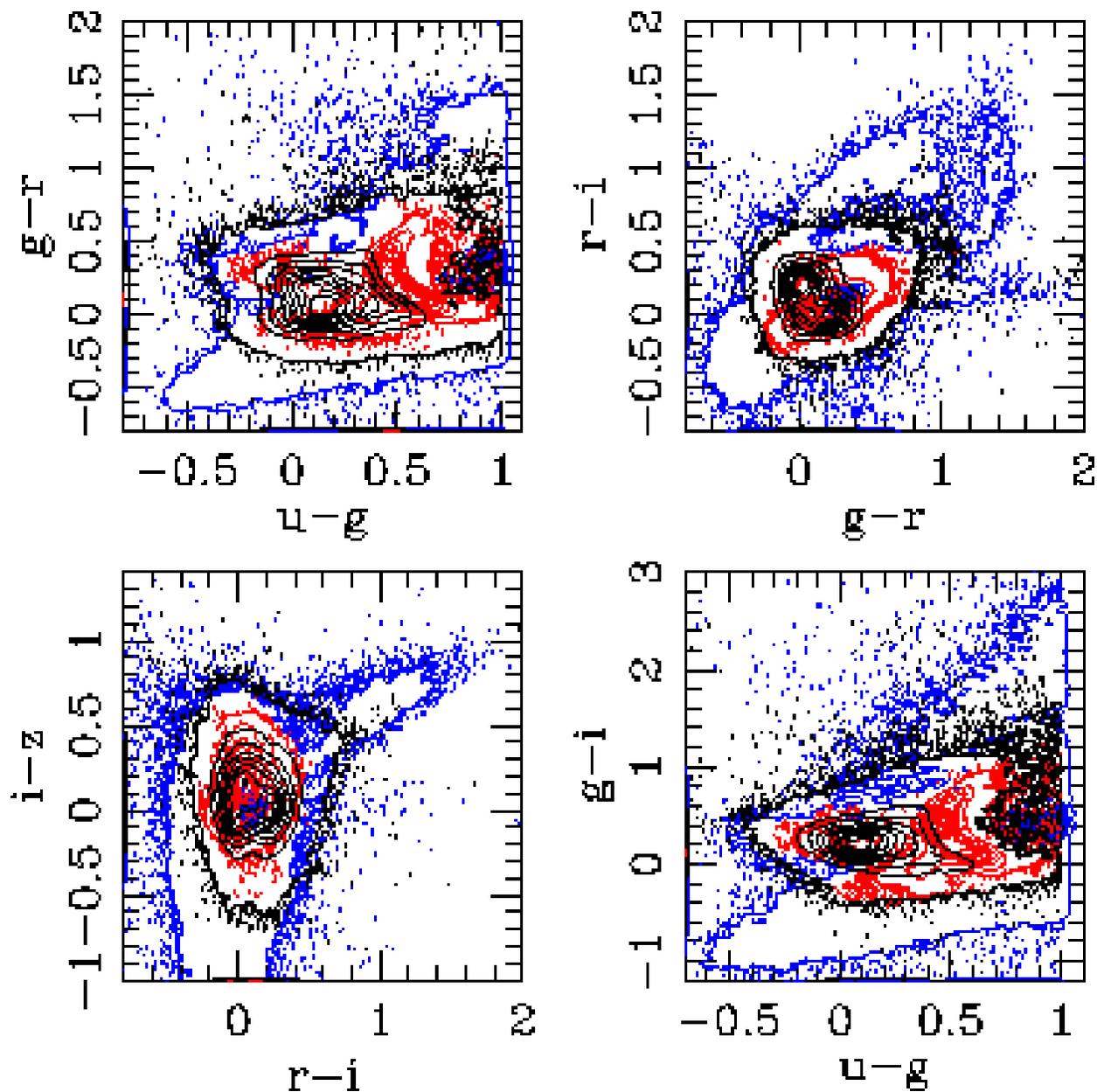}
\caption{Color-color distribution of the 831,600 initial unresolved UVX
sources.  Blue dots and contours are those objects classified as
stars.  Black dots and contours are objects classified as quasars.
Red dots and contours are objects initially classified as quasars but
were rejected by our cut on the stellar density.  Contours are a
fraction of the peak in each class.
\label{fig:fig2}}
\end{figure}

The left panel of Figure~\ref{fig:fig1} shows the log of the
KDE-computed quasar density vs.\ star density for those objects
classified by the NBC algorithm as quasars.  Objects classified as
stars (not shown) populate the upper-left part of the diagram.  Also
evident in this panel is an island of objects (upper right) with
similar quasar and star densities.  Analysis of the color distribution
(Fig.~\ref{fig:fig2}) of these objects suggests that they are stellar
contaminants; they can be excised with a simple cut on stellar density
(dashed line in left panel of Fig.~\ref{fig:fig1}).  Thus, for the
final catalog, NBC-classified quasars were rejected if the stellar
density exceeded $0.01$.  The right panel of Figure~\ref{fig:fig1}
plots the log of the ratio of the quasar density to star density.
Larger values indicate greater quasar probability.  Objects classified
as stars by NBC, which would occupy the region beyond the left extent
of the plot, have already been removed.  Plotted are both those
objects initially classified as quasars by the NBC algorithm ({\em
dotted line}) and those objects that also pass the stellar density cut
shown by the dashed line in the left-hand panel ({\em solid line}).
Our cut in stellar density is seen to remove objects roughly starting
at the minimum between the quasar and the residual star distributions.
The color distribution of the final classification scheme is shown in
Figure~\ref{fig:fig2}.

\subsection{A Note Regarding Errors}

The reader will notice in the above description that there is no
mention of the photometric errors of the objects; this is because we
do not use them explicitly.  However, we do make implicit use of the
errors in the sense that they are ``in the model''.  That is, when we
ask what the relative quasar/star likelihoods are for a given object,
the answer automatically takes into account the smearing of the color
distribution in the training sets due to photometric errors.

If the magnitude (and thus error) distributions of the training sets
were similar, this process is arguably an appropriate manner in which
to handle errors.  That is because we are asking how likely it is for
an object to be scattered out of the stellar locus, given the
distribution of all stars (which includes the photometric errors)
rather than asking whether an object could conceivably be ``pushed
back'' onto the stellar locus given the individual errors of the
object.  There is a difference between the two since a quasar that is
much redder than the stellar locus (in $u-g$) is necessarily much
fainter in $u$ than a star on the stellar locus with the same $g-r$
color but a bluer $u-g$ color.  That is, it is much more likely that
including the errors of a quasar will cause it to be consistent with
the stellar locus than it is for the errors of a star in the locus to
move it out to the location of a quasar with the same $g-r$ color; see
\markcite{rfn+02}{Richards} {et~al.} (2002) for further discussion of
this issue as it affects the SDSS's selection of quasars.

This method of error handling clearly ignores the fact that the errors
are a function of magnitude and the fact that our quasar training set
has fewer faint objects relative to the stars training
set.\footnote{Though the quasar training set, being based on the
\markcite{sfh+03}{Schneider} {et~al.} (2003) SDSS-DR1 quasar catalog,
{\em does} contain $z<2.2$ objects as faint as $g=22$.}  As we try to
push our selection method to fainter magnitudes, accounting for the
magnitude dependence of the errors will become important
(\S~\ref{sec:future}), but for our current limit of $g=21$, the
typical error on the $u-g$ color at the faint limit is only $\sim 0.1$
mag, which is not enough to adversely affect our selection method
especially since the bandwidth is $0.15$ mag.  In the future we hope
to perform a {\em weighted} KDE analysis, which will allow one to
attach a weight (such as the inverse variance) to every point.

\section{The Quasar Catalog}

\begin{figure}[p]
\epsscale{1.0}
\plotone{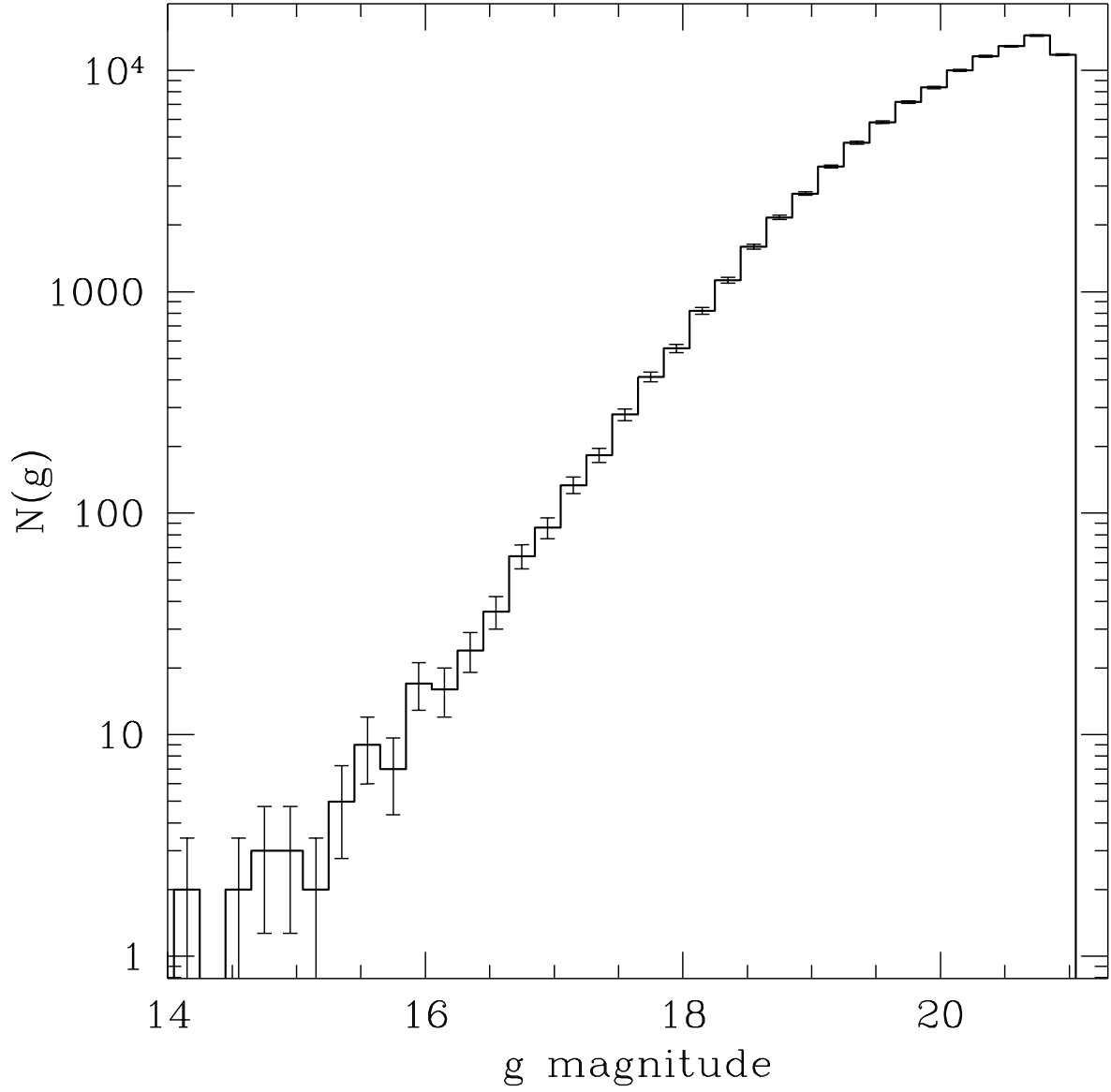}
\caption{$g$ magnitude distribution of the 100,563 objects in the
catalog.  The bin size is 0.2 magnitudes.  Error bars are 1$\sigma$
Poisson errors.
\label{fig:fig3}}
\end{figure}

\begin{figure}[p]
\epsscale{1.0}
\plotone{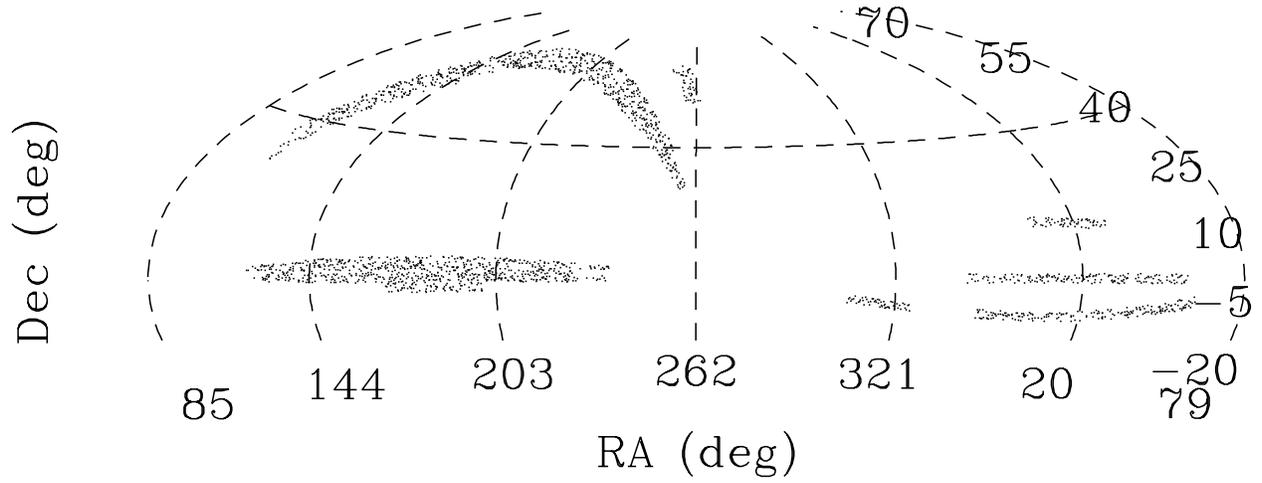}
\caption{Spatial distribution of quasar candidates in an aitoff
projection.  For the sake of clarity, only one in every ten candidates
is shown.
\label{fig:fig4}}
\end{figure}

After application of the NBC algorithm to identify quasars and further
cleaning of this sample by rejecting objects with large KDE stellar
probabilities we are left with $100,563$ quasar candidates that define
this catalog.\footnote{The catalog excludes 17 objects (from run 2206)
that were originally in the test set, but were found to be duplicate
objects resulting from a problem resolving the overlap of 2
(misaligned) strips of data.  This problem is now fixed in the
database.  In addition to the removal of these duplicate objects,
fixing this database problem revealed 125 additional UVX objects (all
from run 2206) that were {\em not} included in the test set from which
we selected our quasar candidates.}  These next sections describe the
completeness and efficiency of the catalog along with matching to
other catalogs.  Table~\ref{tab:tab1} is the catalog itself.  (The
complete version of this table is in the electronic edition of the
Journal; the printed edition contains only a sample.)
Table~\ref{tab:tab2} provides a description of each column in
Table~\ref{tab:tab1}.  The catalog is also available at
http://sdss.ncsa.uiuc.edu/qso/nbckde/, where updates will be posted.
Figure~\ref{fig:fig3} shows the $g$ magnitude distribution of all
sources in the catalog, while Figure~\ref{fig:fig4} shows the
distribution of sources on the sky.

\subsection{Completeness}

The completeness of the sample is difficult to quantify since our
selection extends to both brighter and fainter magnitudes than either
of the SDSS and 2dF \markcite{csb+04}({Croom} {et~al.} 2004) quasar surveys.  However, it is
easy to ask what fraction of SDSS-DR1 quasars are recovered that
should be recovered.

Of the 16,713 SDSS-DR1 quasars in the \markcite{sfh+03}{Schneider}
{et~al.} (2003) catalog, 14,592 meet our magnitude, error, and color
selection cuts.  Among those 14,592, 13,574 are actually in the test
set that formed the basis for our catalog.  Most of the 1018
``missing'' objects result from the fact that we are using only that
imaging area that formally belongs to the DR1 release whereas the
\markcite{sfh+03}{Schneider} {et~al.} (2003) catalog included all
quasars found within any ``stripe'' that was part of the DR1
release.\footnote{An object in a scan that lies along the boundary of
the area observed at the time of a data release may be considered as
the "secondary" observation even if the "primary" observation (from
the adjacent, overlapping scan) does not yet exist.  This is because
such classification of multiple observations is predetermined based on
the geometry of the scans.}  Other objects are missing because of
changes to the object parameters that result from using slightly
different versions of the data processing pipelines.  Among those
13,574 SDSS-DR1 quasars that the NBC algorithm could have recovered, it
actually recovered 12,856 or 94.7\%.  We expect roughly 5\% additional
incompleteness as a result of our filtering of objects via their
photometric flags (e.g., those with ``fatal'' errors); see
\markcite{dvb04}{Vanden Berk et al.} (2004) for further discussion of
the completeness of the SDSS quasar survey.

However, we caution that this completeness is only with respect to the
reasonably bright quasars in the \markcite{sfh+03}{Schneider} {et~al.} (2003) catalog and that we
fully expect that 1) the catalog will be more incomplete with fainter
magnitudes and that 2) the incompleteness of the whole catalog will be
also be a function of redshift and color.  In particular, the fact
that we do not include magnitude as an explicit parameter in our
selection algorithm (other than limiting the magnitude ranges) and the
fact that the colors of stars appear to be a stronger function of
magnitude than the colors of quasars, means that there are regions of
color-space where we are likely to be more incomplete as a result of
our desire to be as efficient as possible.  Utilization of the
magnitudes (see \S~5) in future applications of the algorithm should
improve the completeness in such regions.

We have additionally tested the completeness of the algorithm using
simulations.  Application of the algorithm to simulated quasar colors
constructed similarly to those of \markcite{fan99}{Fan} (1999) reveal
that the algorithm is generally at least 95\% complete between $z=0.2$
and $z=2.0$.  Some additional degree of incompleteness occurs at
$z=0.4$ to $z=0.8$ for the reddest quasars as a result of our
restriction to objects with low stellar likelihood (see \S~3.8).  On
the other hand, the bluest quasars have a 95\% completeness limit that
extends to $z\sim2.4$.  At higher redshift, the completeness drops
rapidly and is difficult to characterize due to the complexity of
accurately simulating quasar spectra blueward of Ly$\alpha$ emission;
furthermore we have restricted this catalog to UVX ($u-g<1.0$)
sources.

\subsection{Efficiency}

To estimate the efficiency of the catalog, we have matched it to three
spectroscopic databases.  First the SDSS-DR1 quasar catalog
\markcite{sfh+03}({Schneider} {et~al.} 2003), which includes only
bonafide quasars; these are objects that constituted part of the
quasar training set and are labeled with ID ``DR1QSO'' in the catalog.
Next we match to the 2QZ NGP catalog \markcite{csb+04}({Croom}
{et~al.} 2004); these objects are labeled with ID ``2QZ'' in the
catalog and include quasars as well as non-quasars.  We also extracted
all ``good''\footnote{Specifically using a database query on the
``SpecPhoto'' database with ``(zConf $>0.95$ AND specClass in
(1,2,3,4,6) AND zStatus in (3,4,6,7,9,11,12)''.} spectroscopic IDs
from the SDSS-DR2 \markcite{aaa+04}({Abazajian et al.} 2004) database
and matched them to our quasar candidate catalog.  These ``DR2''
objects are so labeled and include quasars as well as non-quasars.
They are not meant to be a complete sample of all DR2 identifications,
just those that with identifications that we can be reasonably certain
are correct without having to look at the spectra by eye.  Users
desiring a more complete sample may wish to perform a less restrictive
matching.

Matching to these three catalogs was done in series in the order
given, such that an object will only match the first occurrence.  For
example, any object that matched a SDSS-DR1 quasar was not allowed to
match any other catalog.  In the future, matching against the faint
quasars found in the recently combined SDSS/2dF quasar survey (whose
goal is to discover 10,000 quasars to $g=21.85$ using SDSS imaging and
the 2dF spectrograph; publication in preparation) will allow for
better characterization of the faint end of the sample.

\begin{figure}[p]
\epsscale{1.0}
\plotone{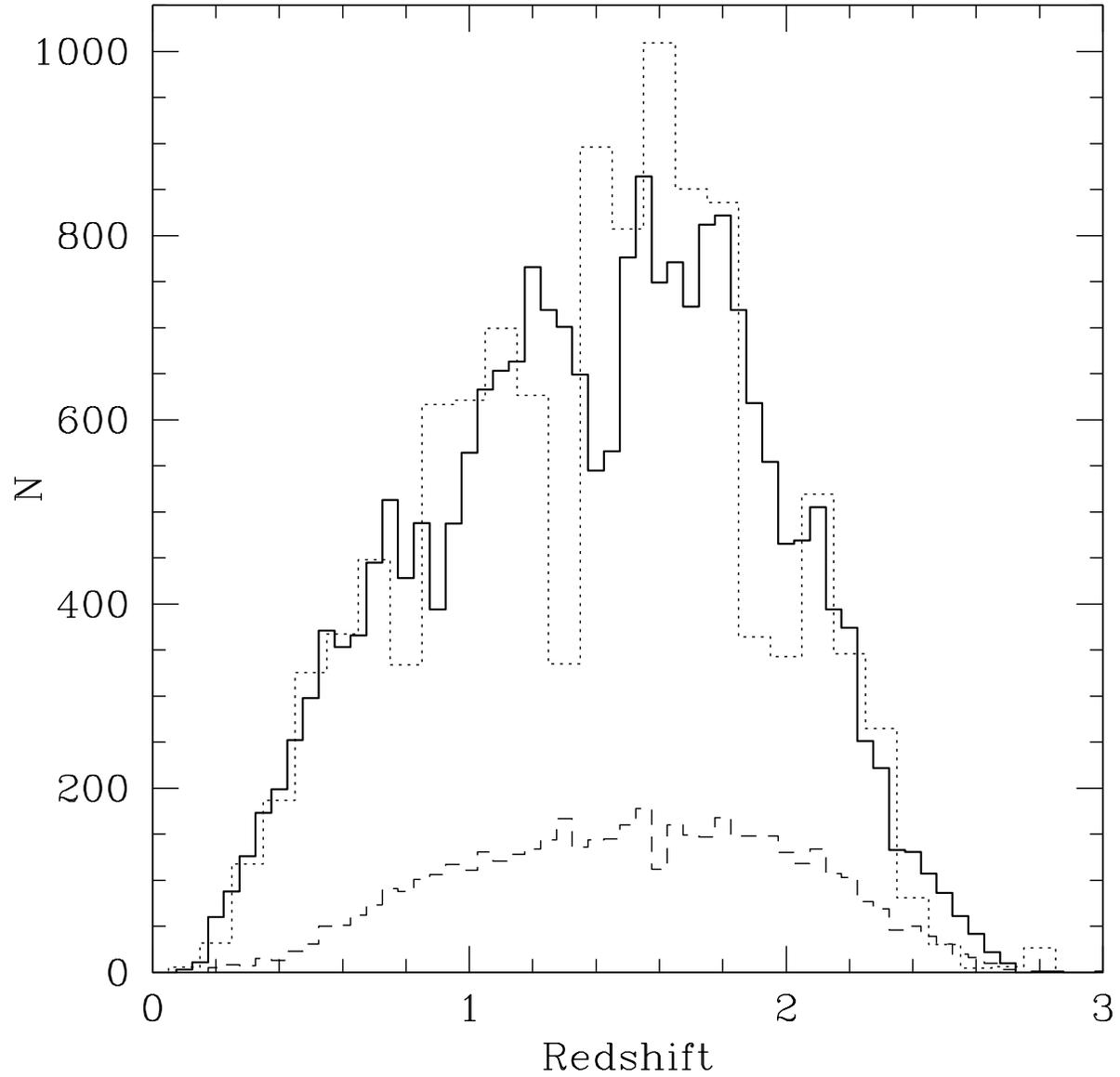}
\caption{Distribution of spectroscopic redshifts for confirmed quasars
in the sample ({\em solid line}) and 2QZ quasars in the sample ({\em
dashed line}).  The {\em dotted} line shows the photometric redshift
distribution of the spectroscopically confirmed quasars.
\label{fig:fig5}}
\end{figure}

In all there were 22,737 matches to spectroscopically confirmed
objects.  A total of 22,191, or 97.6\%, were confirmed to be quasars;
Figure~\ref{fig:fig5} shows the distribution of {\em spectroscopic}
redshifts ({\em solid} line) along with the redshift distribution of
re-discovered 2QZ quasars ({\em dashed line}; \markcite{csb+04}{Croom}
{et~al.} 2004).  Note that some redshifts are intrinsically harder for
the SDSS's automatic spectroscopic identification program to handle.
Our restriction to the most secure identifications when matching to
the DR2 database causes a loss of quasars at certain redshifts (e.g.,
$z\sim0.9, 1.4,\,{\rm and}\;2.0$), which is plainly evident in
Figure~\ref{fig:fig5}.  The distribution of {\em photometric}
redshifts (\S~4.3) for all of the confirmed quasars is given by the
{\em dotted} line in Figure~\ref{fig:fig5}.

The majority of the non-quasars appear to be relatively cool white
dwarfs (see \S~\ref{sec:propmo} below).  Figure~\ref{fig:fig6} shows
the efficiency as a function of $g$ magnitude for the above 22,737
spectroscopic identifications.  Comparison with Figure~\ref{fig:fig3}
reveals that the brightest magnitudes have very few objects in each
bin --- making these fractions less reliable, as indicated by the
error bars in Figure~\ref{fig:fig6}.  Applying the fractions as a
function of magnitude to the overall magnitude distribution of the
catalog, we expect that the overall efficiency of the catalog will be
roughly 95.0\%, yielding 95,502 quasars in all.

It is difficult to extrapolate the efficiency for these confirmed
objects to the entire sample since the selection algorithms of the
three catalogs to which we matched are obviously different from that
herein and we might, for example, be preferentially lacking spectra of
non-quasars.  However, the color distribution of those objects with
matches appears to span the space occupied by the catalog as a whole.
Thus, we fully expect the catalog to be more than 90\% efficient.  In
addition, some objects that were spectroscopically confirmed as
galaxies or narrow emission line galaxies (NELGs) may indeed prove to
be AGN upon close examination.  Similarly, close inspection may reveal
that some of the objects with cool white dwarf colors are actually BL
Lacs.

Contrasting with our estimated 95\% efficiency is that which would be
achieved by making a simple color-cut.  For example, the UVX color-cut
used by \markcite{sg83}{Schmidt} \& {Green} (1983) corresponds roughly to $u-g<0.6$ in the SDSS
photometric system.  There are 97,035 objects with $u-g<0.6$ in the
NBC catalog, whereas the input to our algorithm contains 139,161 such
objects.  If we make the extreme assumption that 95\% of the UVX
objects in our catalog are indeed quasars and that the excess in the
input catalog consists of only contaminants, then this color cut would
yield an efficiency (quasars:quasar candidates) of only 66.2\%.  This
example gives a lower-limit to the efficiency that one can expect for
a reasonably complete sample of UVX quasars; further color-cuts could
obviously be used to improve this efficiency.

\subsection{Photometric Redshifts}

For each object in the catalog, we also report photometric redshifts
that were determined via the method described in \markcite{wrs+04}{Weinstein et al.} (2004).  This
algorithm minimizes the difference between the measured colors of each
object and the median colors of quasars as a function of redshift.

We used the colors of UVX, $u$-detected DR1, point-source quasars with
spectroscopic redshifts as our color-redshift template, but using the
entire DR1 quasar catalog produces similar results.  For each object
we list the most likely photometric redshift\footnote{The precision on
the photometric redshifts is not reflective of the actual accuracy, it
is merely an artifact of our choice of quantized bin centers.}, a
redshift range, and the probability that the redshift is within that
range; see \markcite{wrs+04}{Weinstein et al.} (2004) for more details.

The left panel of Figure~\ref{fig:fig7} shows the spectroscopic versus
photometric redshifts of the 22,191 confirmed quasars in the catalog,
revealing those redshifts where the algorithm has the largest error
rate (either due to degeneracy between distinct redshifts or smearing
of nearby redshifts).  However, one can see from the highly
zero-peaked distribution in the right panel that, overall, the quasar
photo-$z$ algorithm performs quite well, with 19,086 (86.0\%) of the
redshifts being correct to within $\pm0.3$ and 14,371 (64.8\%) to
within $\pm0.1$.  Figure~\ref{fig:fig8} shows the accuracy of the
photometric redshifts as a function of redshift (both spectroscopic
and photometric, {\em left panel}) and $g$ magnitude ({\em right
panel}) for $\Delta z\pm0.3$ and $\Delta z\pm0.1$

The photo-$z$ code also gives a probability of an object being in a
given redshift range (where the size of that range can vary
considerably).  Figure~\ref{fig:fig9} plots the estimated
probability of the photometric redshift being in the given range
versus the actual fraction of those objects with accurate photometric
redshifts --- demonstrating that these probabilities are accurate in
the ensemble average.  Judicious use of the predicted redshifts, the
range given, and the probability of the object having a redshift in
that range allows these photometric redshift estimates to be very
useful for a number of science applications.

\subsection{Matching to Other Catalogs}

Although the estimated efficiency of the algorithm that produces our
catalog is already quite high, it is possible to make use of other
data to improve our efficiency.  For example, objects that match to
radio and/or X-ray sources are that much more likely to be quasars,
while objects with large proper motions are less likely to be quasars.
Thus, included in the catalog are matches to radio, X-ray, and proper
motion catalogs as discussed below.  These matches are primarily for
the purpose of assessing the quasar likelihood of these objects; the
user should refer to the original catalogs for further information.

\subsubsection{FIRST and {\em Spitzer}-FLS VLA}

We have matched the entire catalog to the FIRST \markcite{bwh95}({Becker}, {White}, \& {Helfand} 1995) VLA
20\,cm catalog.  Objects within $1\farcs5$ are considered a match ---
the same radius used for the SDSS's target selection algorithm.
Column 31 of Table~1 indicates the peak 20\,cm flux densities (in mJy)
for those quasars with FIRST matches.  Entries of ``$-1$'' indicate no
radio detection (or no coverage of that position).

In addition, quasar candidates within the {\em Spitzer} First Look
Survey\footnote{http://ssc.spitzer.caltech.edu/fls/} area have been
matched to the deep 20\,cm VLA catalog of \markcite{ccy+03}{Condon} {et~al.} (2003), which goes
approximately 10 times deeper than FIRST in this region of sky.  Those
objects in the catalog that match to an object in the \markcite{ccy+03}{Condon} {et~al.} (2003)
catalog within $1\farcs5$ have their {\em integrated} 20\,cm fluxes
tabulated in Column~31 of Table~1.  Objects that match both radio
catalogs have only their FIRST data reported (as discerned by their
$\gtrsim1$\,mJy flux densities).

In all we catalog 2533 radio detections.  The apparent low fraction of
radio-detected sources should not be taken as an indication that the
fraction of quasars that are radio loud is lower than the nominal
8--10\% \markcite{imk+02}(e.g., {Ivezi{\'c}} {et~al.} 2002).  Rather
it reflects the fact that the catalog is going much deeper in optical
than FIRST does in the radio.

\subsubsection{ROSAT}

We have cross-correlated the positions of the quasar candidates with
the X-ray sources listed in the Bright and Faint Source catalogs of
the ROSAT All-Sky Survey (RASS; \markcite{vab+99,vab+00}{Voges}
{et~al.} 1999, 2000).  Positional accuracies for RASS X-ray sources
vary with count rate, but typically have an uncertainty of
$\sim10$--$30\arcsec$.  Among the SDSS quasar candidates presented here,
there are 1304 cases whose optical positions fall within $30\arcsec$
of a RASS X-ray source; for these sources Column~32 of Table~1 gives
the log of the broadband (0.1--2.4~kev) count rate (counts sec$^{-1}$)
corrected for vignetting.  A ``9'' in Column~32 indicates no X-ray
detection.  Since the surface density of our quasar candidates is
about 45.5 deg$^{-2}$ and since there are about 7000 RASS X-ray
sources within the SDSS DR1 imaging region, the expected number of
SDSS quasar candidates superposed on unrelated RASS $30\arcsec$ radius
X-ray error circles is about 69, i.e., about 5\% of the 1304 likely
SDSS/RASS positional matches we tabulate here.  There are 15 cases in
which two SDSS quasar candidates fall within the same 30$\arcsec$
radius RASS error circle, making their association with the X-ray
source especially ambiguous; the catalog numbers of these ambiguous
candidates are: 769/772, 46119/46120, 49117/49123, 50252/50253,
50751/50756, 51095/51096, 70101/70105, 76539/76540, 79692/79701,
81782/81783, 85386/85387, 86170/86171, 92237/92240, 93927/93934, and
99321/99322.

\subsubsection{USNO-B Proper Motion}
\label{sec:propmo}

Objects with large proper motions are more likely to be stars than
quasars.  We have matched the quasar candidates to the improved
USNO-B+SDSS proper motions tabulated by \markcite{mun04}{Munn} {et~al.} (2004), which is 90\%
complete to $g\sim19.7$.  We chose to restrict ourselves to the most
reliable proper motions, and thus require 1) a one-to-one match
between the SDSS and USNO-B catalogs, 2) that the proper motion rms
fit residual be less than 550 mas in both right ascension and
declination, 3) that the SDSS object be detected in at least four
epochs, and 4) that the nearest neighbor (to $g<22$) be more than
$10\arcsec$ away (to avoid blended objects on the Schmidt plates from
which USNO-B was created, which could lead to false high proper
motions).

This matching results in 41,241 ``reliable'' proper motion
measurements.  Since quasars will have measured ``proper motions''
comparable to the typical errors in the proper motions, we need to
impose a limit on the proper motion to identify objects that are most
likely to be stars.  We find that 99.5\% of spectroscopically
confirmed quasars in our sample have proper motions less than 20 mas
year$^{-1}$.  There are 799 objects with proper motions $\ge20$ mas
year$^{-1}$ in the catalog, most of which are likely to be stars.

In Figure~\ref{fig:fig10} we show the color distribution of the
confirmed quasars ({\em black}) and confirmed non-quasars or large
proper motion objects ({\em red}).  Most of the non-quasars are cool
white dwarfs with colors very similar to real quasars and are thus
difficult to exclude.  Those desiring the most efficient samples
possible may wish to exclude this color region (especially for bright
objects).  Note, however, that the contours are given as a fraction of
the peak in each category and the overall level of contamination is
small.

\section{Future Work \label{sec:future}}

Although the selection algorithm from which this quasar catalog was
derived is very efficient and complete with respect to unresolved UVX
quasars, we can envision modifications that would improve the
algorithm --- especially with regard to high-redshift and extended
quasars.

We currently limit the test set for this catalog to UVX-selected
objects in part because of a lack of a sufficiently large number of
high-$z$ quasars from which to train the algorithm.  That is, since
the density of $z>3$ DR1 quasars is small, the algorithm would tend to
identify any such objects as stars.  However, we are hopeful that in
the future we can use simulated quasars as the quasar training set,
which will overcome the relative underdensity of high-$z$ objects and
should allow for efficient selection of quasars to $z\sim5.8$ with the
NBC method.

In addition, we currently exclude extended sources even though they
may be AGN since their extended morphology necessarily means that the
host galaxy is contributing a significant amount of light to the
object's colors.  We hope to include such objects in the future by
explicitly including an extended quasar and normal galaxy training
sets.  Furthermore, we intend to use a Bayesian star/galaxy
classification algorithm \markcite{sjd+02}({Scranton} {et~al.} 2002) for future catalogs in order
to reduce the fraction of objects that have misclassified
morphologies.

Another improvement that could be made is to push the selection to
fainter limits.  The SDSS imaging data have a $95\%$ completeness
detection limit of 22.2 in the $g$-band which means that it should be
possible to extend our sample to magnitudes fainter than $g=21$.

Similarly we might make better use of the magnitudes in the selection
algorithm itself.  Currently our algorithm makes use of only the four
unique colors that can be derived from the five SDSS magnitudes.
Since errors and metallicity make the width and location of the
stellar locus a function of magnitude, one would like to include a
magnitude in the selection algorithm.  Doing so is a complicated
matter since magnitudes and colors are not distributed similarly and
thus have different ``metrics''.  However, a possible solution for the
future is to use the algorithm on the five SDSS magnitudes rather than
the four SDSS colors.

Our efficiency would also be improved if we were able to include
properties such as radio- and X-ray detections and lack of proper
motion into our algorithm rather than making use of that information
after the fact.

Finally, since the density of stars is clearly a function of Galactic
position, it would be reasonable to make use of this information in
the classification of quasars.  One possibility is to make the stellar
prior a function of Galactic coordinate.

\section{Science}

Here we highlight some of the science applications for which we
envision the catalog being used.  The most obvious of those is the
study of the magnitude distribution of quasars, i.e., their number
counts.

Since 1) the efficiency of our algorithm is so high, 2) the selection
yields a redshift distribution similar to that of the 2QZ survey
\markcite{csb+04}({Croom} {et~al.} 2004), and 3) $g$ and $B_J$ are roughly equivalent, we can
quite easily compare the number counts distribution of our quasar
candidates to that of the 2QZ/6QZ catalog.  Such a comparison is shown
in Figure~\ref{fig:fig11} and shows remarkable agreement, confirming
that the number counts roll over at the faint end and are better fit
by a double power-law than a single power-law (though we prefer not to
use the term ``break'' to describe this behavior).

Further work is needed to properly compare the number counts of
quasars from this catalog to the 2QZ sample and to compute the
luminosity function.  Currently the incompleteness of the catalog and
the contamination of non-quasars are similar in fraction and
fortuitously cancel, possibly making the comparison in
Figure~\ref{fig:fig11} look better than it really is.  However, even
accounting for this, the agreement of the raw catalog to previous work
is a testament to the completeness and efficiency of our method.

In addition to the number counts of quasars, there exists an abundance
of other science that can be done with this catalog.  For example, the
SDSS quasar selection algorithm is forced to reject the brightest
quasars (typically $i=15$, but as faint as $i=16.5$ in the early data)
to avoid cross-talk between the spectroscopic fibers.  As a result
there 22 bright quasar candidates in the catalog without matches to
our catalogs of known objects; 13 of these are unknown to
NED\footnote{The NASA/IPAC Extragalactic Database (NED) is operated by
the Jet Propulsion Laboratory, California Institute of Technology,
under contract with the National Aeronautics and Space
Administration.} (objects 2047, 5398, 16881, 20333, 23715, 46200,
50830, 83155, 93643, 95179, 95336, 95341, and 97262).  Spectroscopy of
three of these objects with the ARC 3.5m telescope at Apache Point
Observatory shows that objects 46200 and 50830 are not quasars, while
object 83155 is a $z=0.039$ AGN.  Based on their colors, objects 2047,
5398, and 95341 are the mostly likely AGN candidates among the
remaining 10 unknown objects.

An obvious application of this catalog is to find wide-separation
($\theta\gtrsim2\arcsec$) gravitational lens candidates that are
clearly resolved in ground-based data.  Not only is it possible to
find pairs of close quasars, but it is also possible to determine the
likely similarity of their redshifts.  Hennawi et al.\ (2005, in
preparation) discuss such a search.  Preliminary application of our
algorithm to post-DR1 successfully recovers (with ``correct''
redshifts) 3 of the 4 components of the quadruple lens,
SDSS~J1004+4112 (the 4th component appears to be dust reddened;
\markcite{ina+03}{Inada} {et~al.} 2003) and also both components of
Q~0957+561 (\markcite{wcw79}{Walsh}, {Carswell}, \& {Weymann} 1979),
again with redshifts accurate to within the errors.


One can also use the catalog to measure the amplification bias of
quasars.  That is, to what extent are quasars magnified (but not split
into multiple images) by foreground galaxies?  Such studies require
cross-correlation of the largest possible samples of foreground
galaxies and background quasars.  The efficiency of the algorithm is
sufficient for such applications.  Furthermore, since we give the
probability of the photometric redshifts in addition to the most
likely value, it is possible to exclude quasars that may not be
background sources to samples of foreground SDSS galaxies.

Finally, we emphasize that the expected density of UVX quasars in this
catalog (45.5 deg$^{-2}$) is substantially larger than the density of
6.95 deg$^{-2}$ for similar objects from \markcite{sfh+03}{Schneider} {et~al.} (2003), so this
sample will be very powerful for investigations of quasar-quasar and
quasar-galaxy clustering.

\acknowledgements

Funding for the creation and distribution of the SDSS Archive has been
provided by the Alfred P. Sloan Foundation, the Participating
Institutions, the National Aeronautics and Space Administration, the
National Science Foundation, the U.S. Department of Energy, the
Japanese Monbukagakusho, and the Max Planck Society. The SDSS Web site
is http://www.sdss.org/.  The SDSS is managed by the Astrophysical
Research Consortium (ARC) for the Participating Institutions. The
Participating Institutions are The University of Chicago, Fermilab,
the Institute for Advanced Study, the Japan Participation Group, The
Johns Hopkins University, Los Alamos National Laboratory, the
Max-Planck-Institute for Astronomy (MPIA), the Max-Planck-Institute
for Astrophysics (MPA), New Mexico State University, University of
Pittsburgh, Princeton University, the United States Naval Observatory,
and the University of Washington.  Based also on observations obtained
with the Apache Point Observatory 3.5-meter telescope, which is owned
and operated by the Astrophysical Research Consortium.  This research
has made use of the NASA/IPAC Extragalactic Database (NED) which is
operated by the Jet Propulsion Laboratory, California Institute of
Technology, under contract with the National Aeronautics and Space
Administration.  This work was partially supported by NSF grant AST
03-07582.

\clearpage



\clearpage

\begin{figure}[p]
\epsscale{1.0}
\plotone{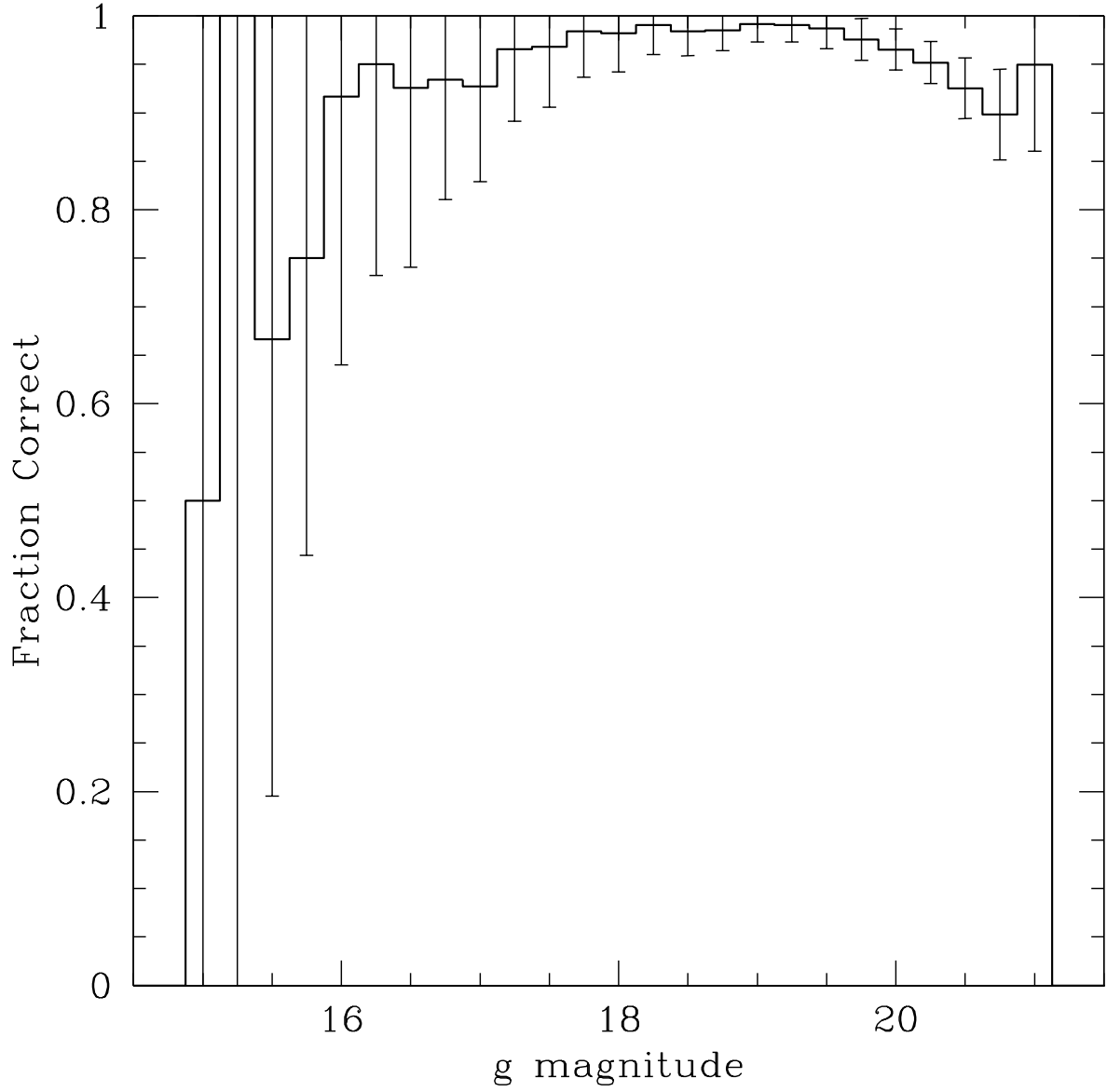}
\caption{Efficiency of the target selection as a function of $g$
magnitude for 22743 spectroscopically confirmed objects.  Error bars
are 1$\sigma$ Poisson errors.
\label{fig:fig6}}
\end{figure}

\begin{figure}[p]
\epsscale{1.0}
\plottwo{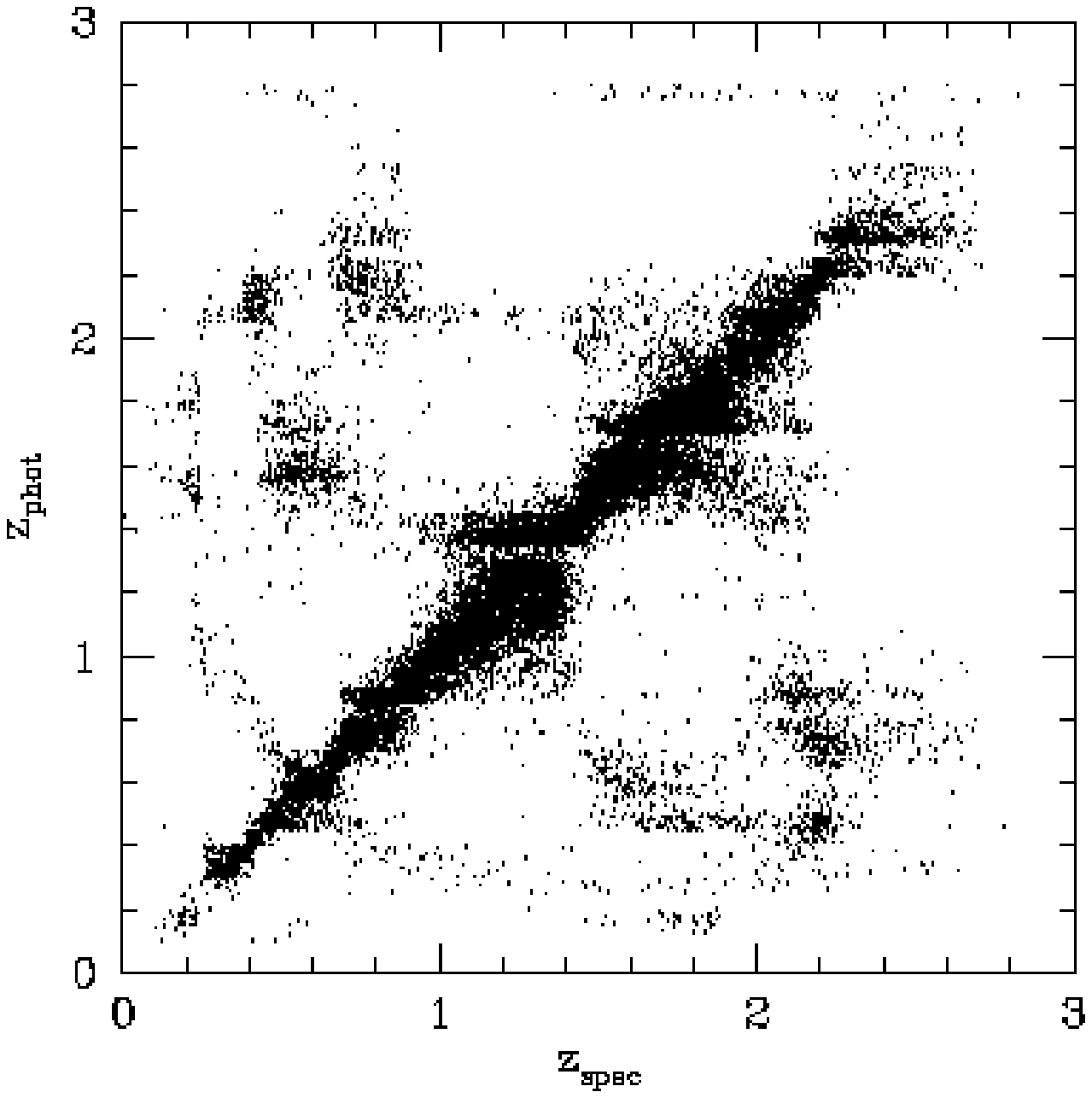}{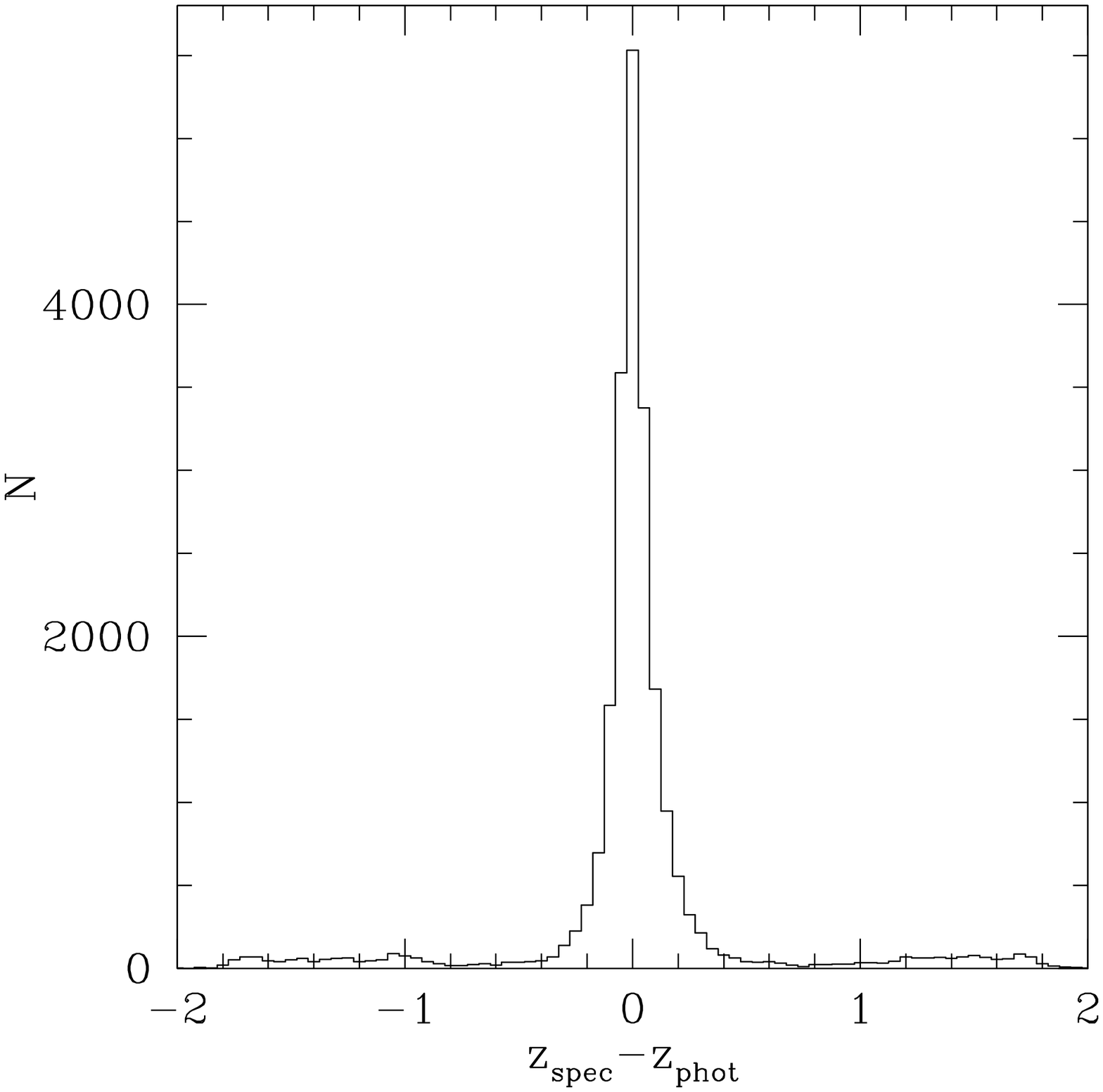}
\caption{{\em Left:} Spectroscopic vs.\ photometric redshifts.  {\em
Right:} Histogram of the difference between spectroscopic and
photometric redshifts.
\label{fig:fig7}}
\end{figure}

\begin{figure}[p]
\epsscale{1.0}
\plottwo{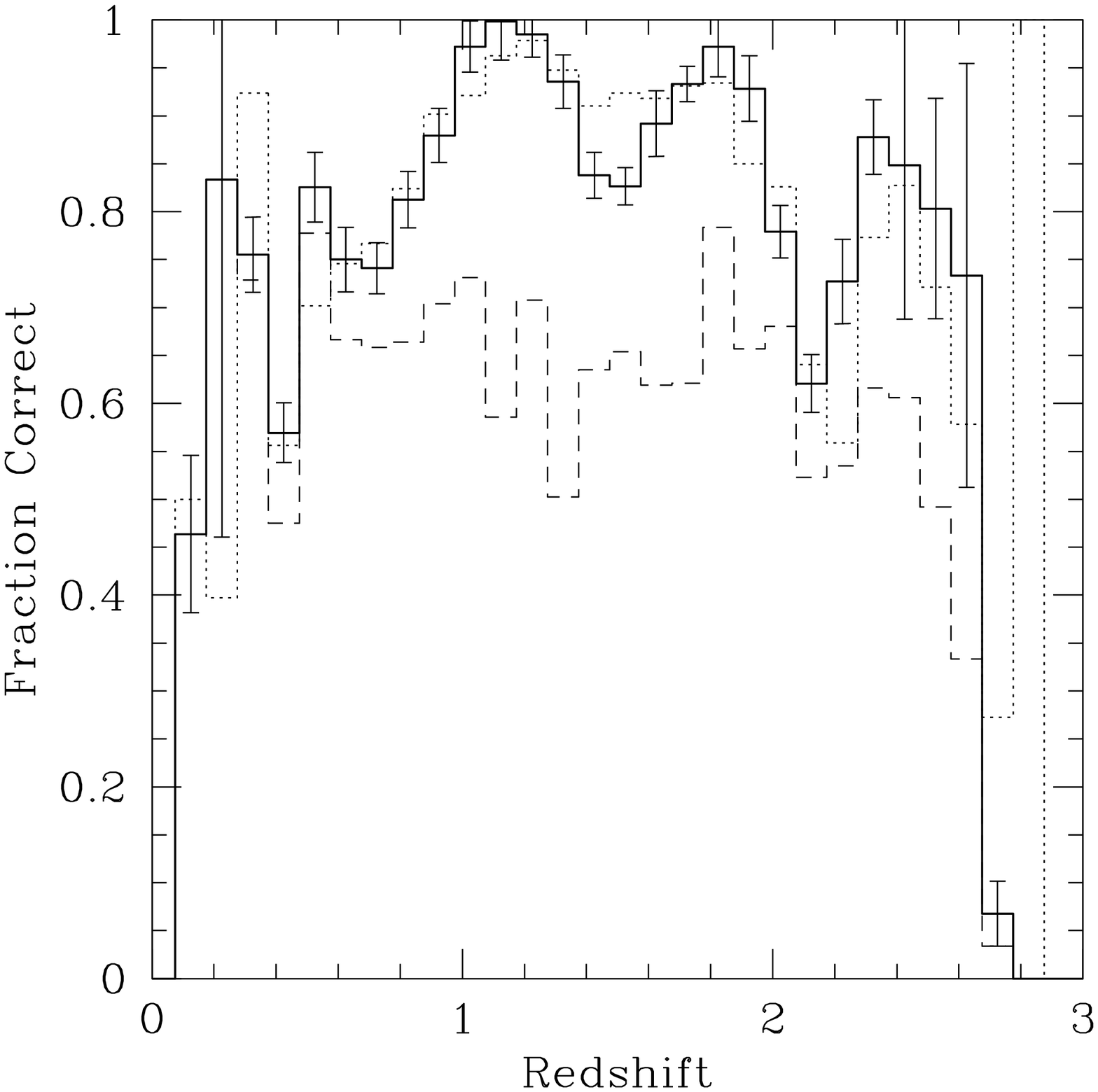}{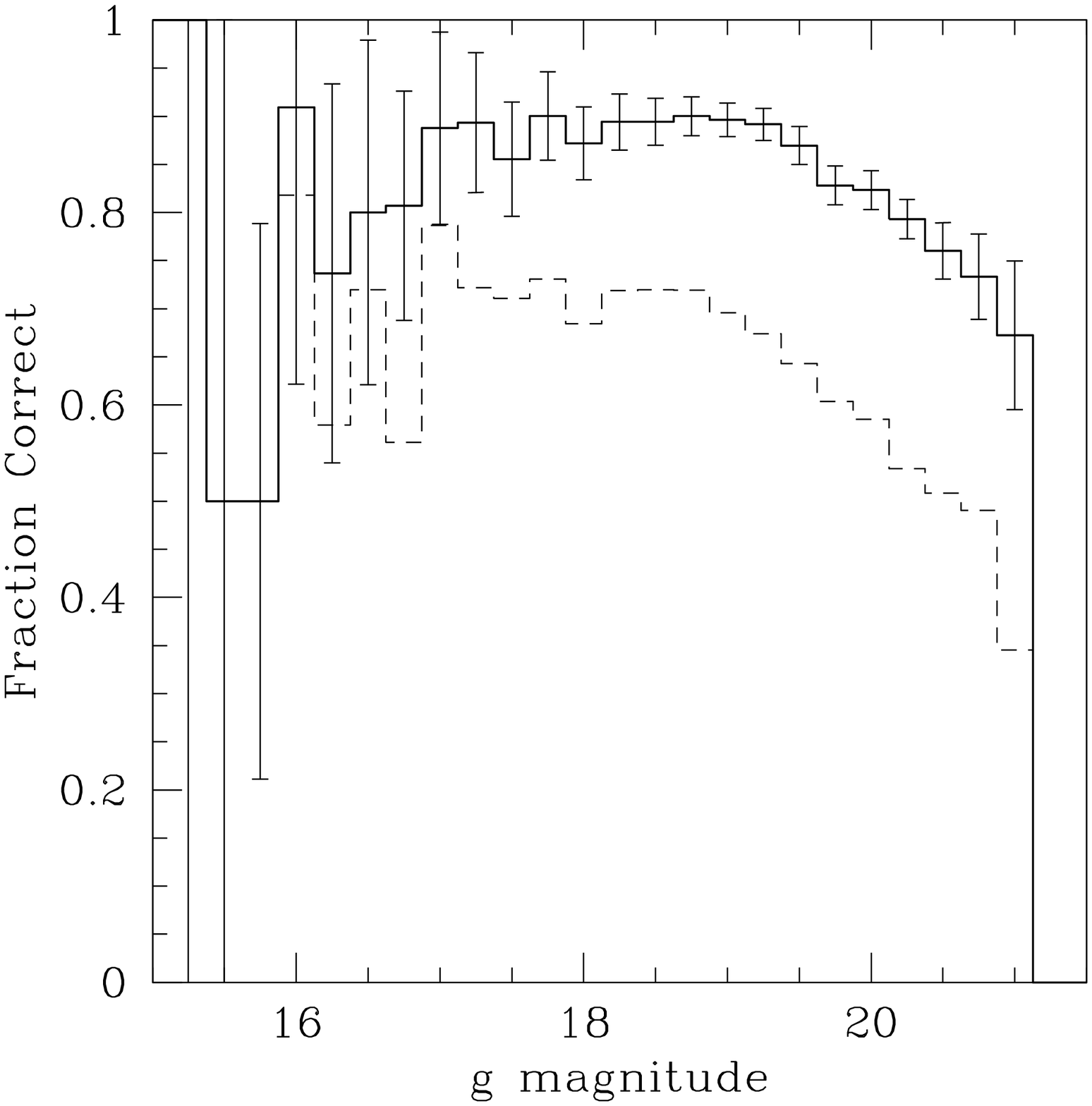}
\caption{{\em Left:} Fraction of correct photometric redshifts as a
function of {\em spectroscopic} redshift, $z_{\rm spec}$, as a
function of redshift ({\em solid}: $\Delta z\pm0.3$, error bars are
1$\sigma$ Poisson errors; {\em dashed $\Delta z\pm0.1$}).  The {\em
dotted line} shows the fraction of correct photometric redshifts as a
function of {\em photometric} redshift, $z_{\rm phot}$, for $\Delta
z\pm0.3$.  {\em Right:} Fraction of correct photometric redshifts as a
function of $g$; error bars are 1$\sigma$ Poisson errors.
\label{fig:fig8}}
\end{figure}

\begin{figure}[p]
\epsscale{1.0}
\plotone{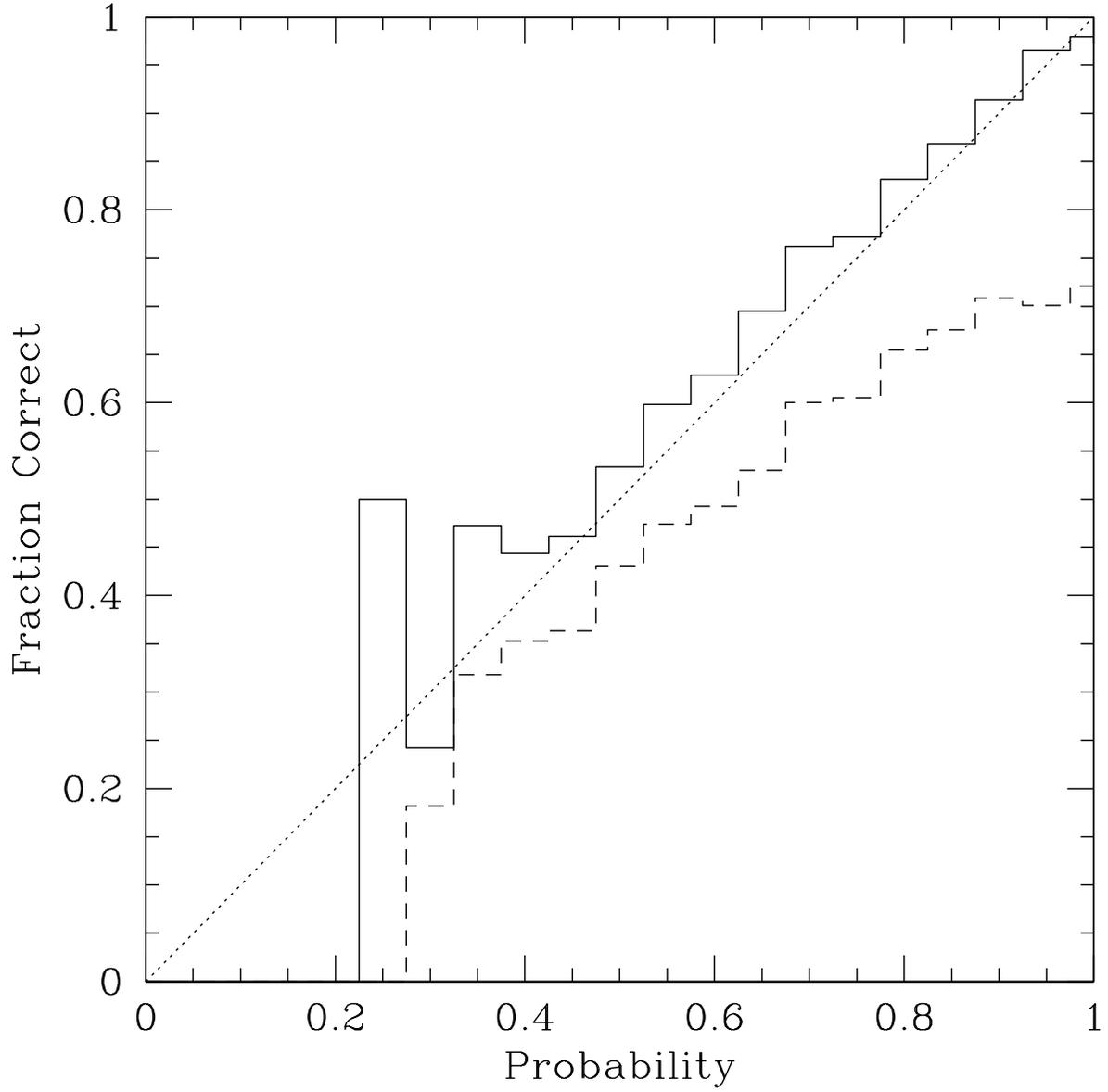}
\caption{Actual fraction with correct redshift as a function of the
quoted probability that the redshift is correct ({\em solid}: $\Delta
z\pm0.3$; {\em dashed $\Delta z\pm0.1$}).
\label{fig:fig9}}
\end{figure}

\begin{figure}[p]
\epsscale{1.0}
\plotone{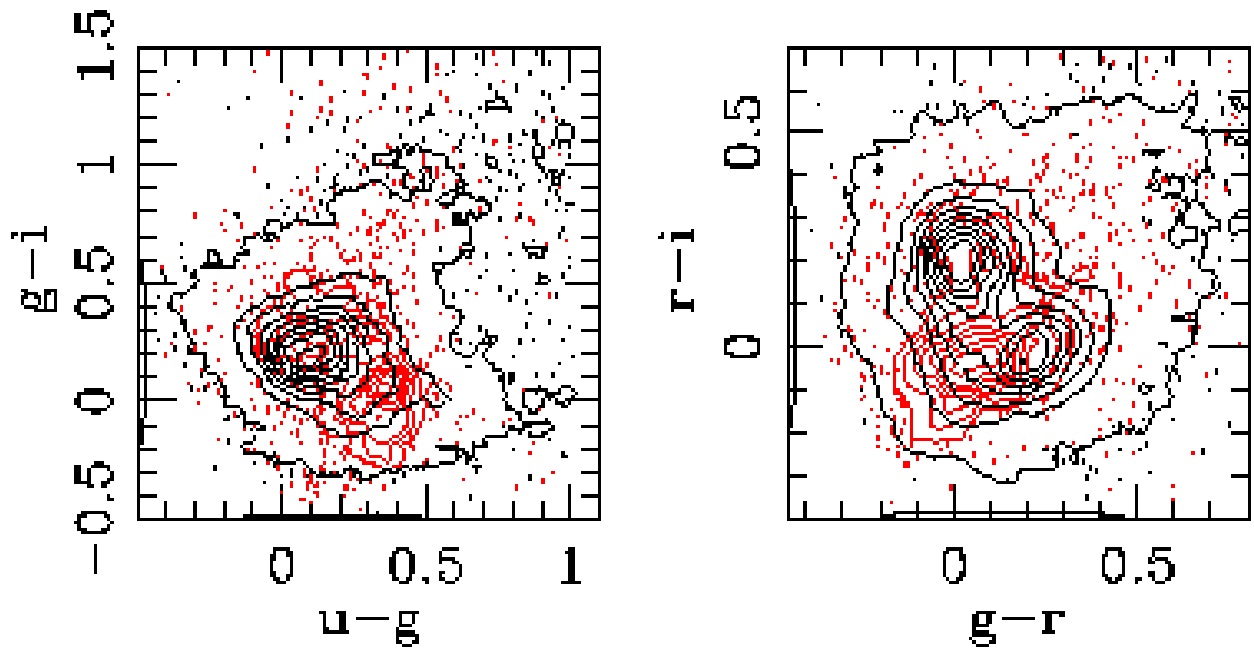}
\caption{Color distribution of confirmed quasars ({\em black}) and
confirmed non-quasars (mostly white dwarfs) or large proper motion
objects ({\em red}).  Note the small region of parameter space
occupied by residual white dwarfs.  However, their inclusion does not
significantly affect the distribution of quasar candidates and thus
are a small contaminant overall (the contours are given as a fraction
of the peak in each class).
\label{fig:fig10}}
\end{figure}

\begin{figure}[p]
\epsscale{1.0}
\plotone{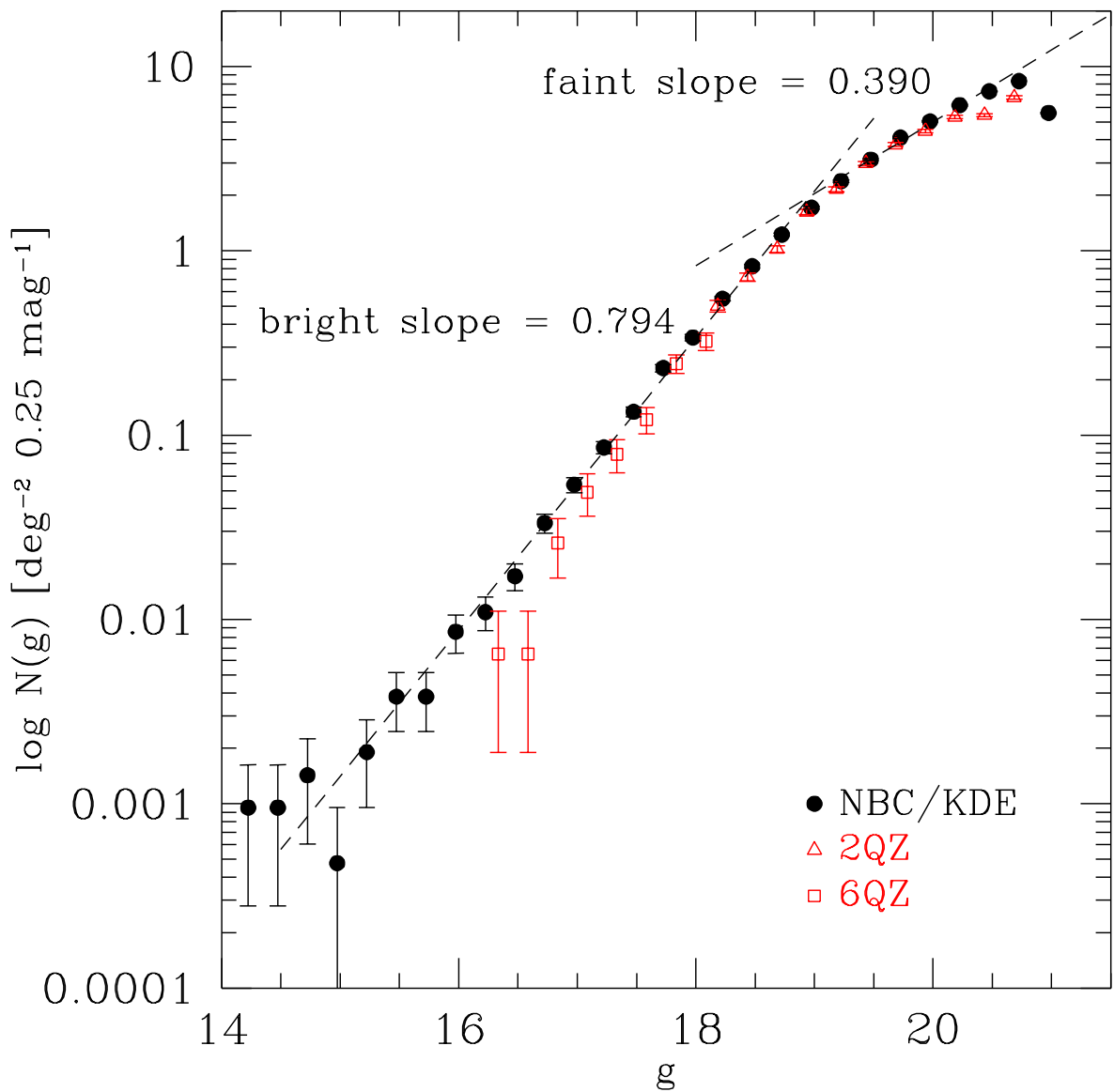}
\caption{Number counts of NBC quasar candidates (after removal of
confirmed contaminants and high proper motion objects), as compared to
the $z<2.2$ number counts from the 2QZ and 6QZ surveys.
\label{fig:fig11}}
\end{figure} 

\clearpage

\begin{deluxetable}{llrrlrrrrrrr}
\tabletypesize{\footnotesize}
\rotate
\tablewidth{0pt}
\tablecaption{NBC Quasar Candidate Catalog\label{tab:tab1}}
\tablehead{
\colhead{Num} &
\colhead{Name} &
\colhead{R.A.} &
\colhead{Decl.} &
\colhead{ObjID} &
\colhead{row} &
\colhead{col} &
\colhead{$u$} &
\colhead{$g$} &
\colhead{$r$} &
\colhead{$i$} &
\colhead{$z$} \\
\colhead{} &
\colhead{(SDSS J)} &
\colhead{(deg)} &
\colhead{(deg)} &
\colhead{} &
\colhead{} &
\colhead{} &
\colhead{} &
\colhead{} &
\colhead{} &
\colhead{} &
\colhead{}
}
\startdata
1 & 000001.88$-$094652.1 & 0.0078478 & $-$9.7811413 & 1-1729-21-4-83-116 & 370.57 & 1729.17 & 19.781 & 19.530 & 19.335 & 19.401 & 19.407 \\
2 & 000002.21$-$094956.0 & 0.0092176 & $-$9.8322327 & 1-1729-21-4-83-118 & 389.98 & 1264.98 & 20.396 & 20.281 & 20.296 & 20.209 & 20.152 \\
3 & 000006.53+003055.2 & 0.0272316 & 0.5153435 & 1-3325-20-5-108-117 & 656.47 & 978.59 & 20.405 & 20.459 & 20.336 & 20.100 & 20.076 \\
4 & 000007.58+002943.3 & 0.0316062 & 0.4953686 & 1-3325-20-5-108-131 & 696.30 & 797.03 & 21.085 & 20.440 & 20.471 & 20.336 & 19.958 \\
5 & 000008.13+001634.6 & 0.0339044 & 0.2762998 & 1-2662-20-4-283-149 & 253.50 & 673.27 & 20.240 & 20.201 & 19.949 & 19.498 & 19.194 \\
\enddata
\tablecomments{The complete version of this table is in the electronic
edition of the Journal.  The printed edition contains only a sample.}
\end{deluxetable}

\begin{deluxetable}{lcl}
\tabletypesize{\small}
\tablewidth{0pt}
\tablecaption{NBC Quasar Candidate Catalog Format\label{tab:tab2}}
\tablehead{
\colhead{Column} &
\colhead{Format} &
\colhead{Description}
}
\startdata
1 & I6 & Unique catalog number \\
2 & A18 & Name: SDSS J$hhmmss.ss+ddmmss.s$ (J2000.0) \\
3 & F11.7 & Right ascension in decimal degrees (J2000.0) \\
4 & F11.7 & Declination in decimal degrees (J2000.0) \\
5 & A21 & SDSS Object ID string: skyVersion-run-rerun-camcol-field-id \\
6 & F7.2 & Row position of object in field (pixel) \\
7 & F7.2 & Column position of object in field (pixel) \\
8 & F7.3 & PSF $u$ asinh magnitude (uncorrected for Galactic extinction) \\
9 & F6.3 & PSF $g$ asinh magnitude (uncorrected for Galactic extinction) \\
10 & F6.3 & PSF $r$ asinh magnitude (uncorrected for Galactic extinction) \\
11 & F6.3 & PSF $i$ asinh magnitude (uncorrected for Galactic extinction) \\
12 & F6.3 & PSF $z$ asinh magnitude (uncorrected for Galactic extinction) \\
13 & F6.3 & Error in PSF $u$ asinh magnitude \\
14 & F5.3 & Error in PSF $g$ asinh magnitude \\
15 & F5.3 & Error in PSF $r$ asinh magnitude \\
16 & F5.3 & Error in PSF $i$ asinh magnitude \\
17 & F5.3 & Error in PSF $z$ asinh magnitude \\
18 & F6.3 & Galactic extinction (magnitudes) in $u$ \\
19 & F5.3 & Galactic extinction (magnitudes) in $g$ \\
20 & F5.3 & Galactic extinction (magnitudes) in $r$ \\
21 & F5.3 & Galactic extinction (magnitudes) in $i$ \\
22 & F5.3 & Galactic extinction (magnitudes) in $z$ \\
23 & E11.4 & KDE quasar density \\
24 & E10.4 & KDE star density \\
25 & F6.3 & Photometric redshift \\
26 & F4.2 & Lower limit of photometric redshift range \\
27 & F4.2 & Upper limit of photometric redshift range \\
28 & F5.3 & Photometric redshift range probability \\
29 & A13 & Previous catalog object classification \\
30 & F6.3 & Previous catalog object redshift \\
31 & F9.3 & 20\,cm flux density (mJy) ($-1$ for not detected or not covered) \\
32 & F6.3 & log RASS full-band count rate (9 for not detected or not covered) \\
33 & F6.2 & Proper motion (mas year$^{-1}$) ($-1$ indicates unknown proper motion)
\enddata
\end{deluxetable}

\end{document}